\documentclass[preprint,aps,prd,showpacs,nofootinbib]{revtex4}

\usepackage{graphicx}
\usepackage{amsfonts}
\usepackage{amsmath}
\usepackage{color}

\newcommand{\timestwo}{\!\times\!}

\newcommand*{\chpt}{\raise0.4ex\hbox{$\chi$}PT}
\newcommand*{\schpt}{S\raise0.4ex\hbox{$\chi$}PT}

\newcommand*{\et}{\textit{et al.}}

\newcommand*{\Tr}{\textrm{Tr}}

\newcommand*{\amulo}{\ensuremath{a_\mu^{\rm HLO}}}
\newcommand*{\gtwo}{\ensuremath{g\!-\!2}}

\newcommand{\cL}{\mathcal{L}}
\newcommand{\cM}{\mathcal{M}}

\newcommand{\cO}{\mathcal{O}}

\newcommand{\cV}{\mathcal{V}}

\def\eq#1{Eq.~(\ref{eq:#1})}
\def\eqs#1#2{Eqs.~(\ref{eq:#1}) and (\ref{eq:#2})}

\begin{document}

\bibliographystyle{h-physrev}

\preprint{CU-TP-1157, RBRC-609}

\title{Calculating the hadronic vacuum polarization and leading hadronic contribution 
to the muon anomalous magnetic moment with improved staggered quarks}
\author{C.\ Aubin}
\affiliation{Physics Department, Columbia University, New York,
NY 10027}
\author{T.\ Blum}
\affiliation{Physics Department, University of Connecticut, 
Storrs, CT 06269-3046}
\affiliation{RIKEN BNL Research Center, Brookhaven National Laboratory, Upton, NY, 11973-5000}
\begin{abstract}
We present a lattice calculation of the hadronic vacuum polarization and the lowest-order
hadronic contribution (HLO) to the muon anomalous magnetic moment, $a_\mu = (g-2)/2$, using
$2\!+\!1$ flavors of improved staggered fermions. A precise fit to the
low-$q^2$ region of the vacuum polarization is necessary to accurately
extract the muon \gtwo. To obtain this fit, we use staggered chiral
perturbation theory, including a model to incorporate 
the vector particles as
resonances, and compare these
to polynomial fits to the lattice data. We discuss the
fit results and associated systematic uncertainties, paying
particular attention to the relative contributions of the pions and
vector mesons. 
Using a single lattice spacing ensemble generated by the MILC Collaboration ($a=0.086$ fm), light
quark masses as small as roughly one-tenth the strange quark mass,
and volumes as large as (3.4 fm)$^3$, 
we  find $\amulo = (713 \pm 15) \times 10^{-10}$ and  $(748 \pm 21) \times 10^{-10}$ where the error is statistical only and the two values correspond to linear and quadratic extrapolations in the light quark mass, respectively. Considering various systematic uncertainties not eliminated in this study (including a model of vector resonances used to fit the lattice data and the omission of disconnected quark contractions in the vector-vector correlation function),
we view this as agreement with the current best calculations using the experimental cross section for $e^+e^-$ annihilation to hadrons,
692.4 (5.9) (2.4)$\times 10^{-10}$,
and including the experimental decay rate of the tau lepton to hadrons,
 711.0 (5.0) (0.8)(2.8)$\times 10^{-10}$. We discuss several ways to improve the current lattice calculation. 

\end{abstract}
\pacs{12.38.Gc,  
	  12.39.Fe,
	  13.40.Em, 
	  14.60.Ef, 
	  14.65.Bt 
}

\maketitle

\section{Introduction}\label{sec:intro}

The current Standard Model of particle physics has had tremendous
success in describing fundamental properties of Nature. 
Currently,
precise determinations of physical quantities need to be performed in
order to find discrepancies, if any, between theoretical
predictions and experimental measurements. Many, but not all, quantities agree
quite well between theory and experiment. In those cases where there are discrepancies, there is a need to reduce the errors on theoretical and experimental determinations of said quantities to see if these discrepancies are significant.

An excellent candidate for this improvement is the anomalous magnetic moment
of the muon, $a_\mu= (\gtwo)/2$. The experimental world average is
$a_\mu^{\rm exp} 
= 116\, 592\, 080 (5.4)(3.3)\timestwo 10^{-11}$ 
\cite{Bennett:2006fi,bnl_g2}. The incredible precision of this measurement arises from the recent experiments at Brookhaven National Laboratory \cite{Bennett:2006fi}. This
number deviates from current theoretical predictions by
0.7 to 2.7 standard deviations, 
depending on the method of obtaining the
Standard Model number (see the recent reviews  \cite{g2REVIEW,Davier:2004gb}).

\begin{figure}[ht]
\begin{center}
\includegraphics{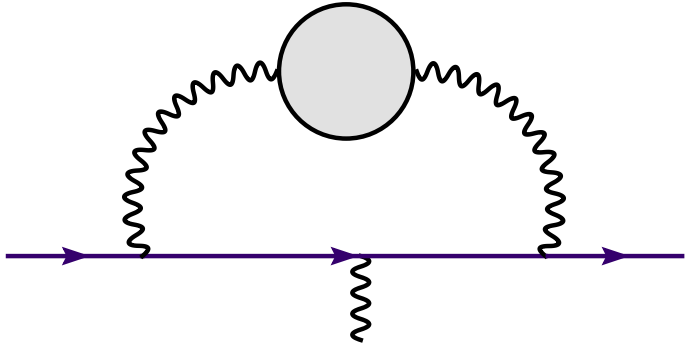}
\caption{The lowest order diagram giving rise to QCD corrections to the muon \gtwo, at $\cO(\alpha^2)$. The blob represents all possible hadronic states.}
\label{fig:QCD_g-2}
\end{center}
\end{figure}

The largest contribution to the error of the theoretical calculation lies
in terms arising from QCD. The electroweak
sector can be reliably calculated using perturbation theory but this
is not the case for QCD. The first QCD contribution comes in at
order ${\cal O}(\alpha^2)$ ($\alpha=e^2/4\pi$ is the fine structure 
constant) shown in Figure~\ref{fig:QCD_g-2} which is the
hadronic contribution to the photon vacuum polarization. The next 
QCD contributions come in at ${\cal O}(\alpha^3)$, the light-by-light contribution, which is currently being studied using lattice techniques \cite{LIGHTBYLIGHT}, model calculations ({\it e.g.},
\cite{Knecht:2001qg,Hayakawa:2001bb,Bijnens:2001cq,Melnikov:2003xd}), and chiral perturbation theory \cite{Ramsey-Musolf:2002cy} (for a recent review, see Ref.~\cite{LightByLightReview}). We focus here on the larger of these two contributions, that coming from the QCD corrections to the photon vacuum polarization. 

Since
perturbation theory is unreliable for such hadronic contributions, for a long time one
has relied on a dispersion relation to relate the
imaginary part of the photon vacuum polarization to the real part
\cite{Bouchiat:1961,Gourdin:1969dm} which is needed in the diagram shown in Figure~\ref{fig:QCD_g-2}, using the experimental $e^+e^-\to {\rm hadrons}$ total cross-section to obtain the imaginary part. Additionally, one can relate the decay $\tau \to$ hadrons  to the $e^+e^-$ cross section in the isospin limit \cite{g2REVIEW,Davier:2004gb} for a part of the energy range. Whether or not one includes the $\tau$ data gives the range of deviations discussed above between theory and experiment. 

A purely theoretical calculation of $\amulo$ is desirable for at least three reasons: first, it is very important to corroborate the dispersive/experimental analysis with an alternative calculation in order to have confidence in the Standard Model  comparison with experiment, and second, these calculations may eventually compete with or surpass the precision and accuracy of the dispersion relation/experiment combination. 
Finally, from a theoretical perspective it is quite interesting.
Such theoretical calculations may in the future provide even more stringent tests between the
Standard Model and experiment, a desirable outcome in the
LHC era.

At present the only way to compute $\amulo$ theoretically in a model-independent way is to apply lattice gauge theory techniques to calculate the hadronic vacuum polarization. This has already been done in the quenched approximation
with domain wall fermions \cite{g-2_1} and improved Wilson fermions \cite{QCDSF}, as well as with dynamical ($2\!+\!1$ light flavors) ``asqtad"
staggered fermions \cite{g-2_2,g-2_3}. The main difficulty is that the largest uncertainties occur in the low-momentum region which gives the dominant contribution to $\amulo$. 
The goal of the current work is to incorporate
staggered chiral perturbation theory (\schpt)
\cite{SCHPT} coupled to photons 
to derive a physically based fitting function to
accurately describe this low-momentum region. Initial results from this fitting technique were presented in  \cite{Lat05g-2}, and to our knowledge are the first chiral perturbation theory results for the 
vacuum polarization presented in the literature.

At this stage we must point out that an expansive and lively debate has recently taken place in the literature, and the lattice QCD community in general, concerning the theoretical validity of the staggered fermion formulation when applied to numbers of quarks different from a multiple of four, as in the present case of 2+1 flavors.
We refer the interested reader to the most recent papers \cite{Shamir:2006nj,Bernard:2006ee,Bernard:2006vv,Bernard:2006zw,Hasenfratz:2006nw,Creutz:2006ys}, and also to the review by Sharpe at Lattice 2006 \cite{SHARPE_LATTICE}. The important point stressed in the review is that staggered fermions are non-local for non-zero lattice spacing, but that the non-locality is a lattice artifact that vanishes in the continuum limit, provided the quark mass is not taken to zero first.  Since we work at fixed lattice spacing, this may be cause for concern. As we shall see, the results presented below do not indicate signs of non-locality, and our future plans include taking the continuum limit when ensembles with
smaller lattice spacing become available\footnote{The MILC collaboration is currently generating such an ensemble.}. We also plan to repeat our calculation with 2+1 flavors of domain wall fermions as a final check.

This paper is outlined as follows. We discuss in 
Sec.~\ref{sec:lat_calc} the basic details of the lattice calculation of the vacuum polarization and how to use this to extract the lowest order hadronic contribution to the muon \gtwo. In Sec.~\ref{sec:schpt_photons} we discuss staggered chiral perturbation theory coupled to photons as external fields, and compute the one-loop pion and kaon contribution to the photon vacuum polarization. We find here that this is not an adequate description of the lattice data, and in Sec.~\ref{sec:vector_schpt} we incorporate into the chiral perturbation theory the lightest vector fields, whose contribution to the vacuum polarization we find dominates over the pions. Sections \ref{sec:fits} and ~\ref{sec:disc} include our discussion of the fits, final results for the muon \gtwo\ for this lattice spacing, and systematic uncertainties in the calculation. We conclude our study in Sec.~\ref{sec:conc}.

\section{Lattice Calculation of the muon \gtwo}\label{sec:lat_calc}

In the continuum, Euclidean and gauge invariance along with the Ward-Takahashi identity allow us to write the vacuum polarization tensor in momentum space as
\begin{equation}\label{eq:WI}
	\Pi_{\mu\nu}(q)  =  (q^2\delta_{\mu\nu} - q_\mu q_\nu)\Pi(q^2)\ .
\end{equation}
We can use $\Pi(q^2)$ as defined above in the space-like region to calculate the lowest order contribution to the muon \gtwo\ from the expressions \cite{g-2_1}
\begin{eqnarray}
	\amulo &=& \left(\frac{\alpha}{\pi}\right)^2
	\int_0^\infty dK^2 f(K^2)\hat\Pi(K^2)\label{eq:amu}\ ,\\
	f(K^2) & = & \frac{m^2_\mu K^2 Z^3 \left(1-K^2 Z\right)}
	{1 + m^2_\mu K^2 Z^2}\label{eq:kernel}\ ,  \\
	Z & = & -\frac{K^2 - \sqrt{K^4 + 4 m^2_\mu K^2}}
	{2 m^2_\mu K^2}\ ,\\
	\hat \Pi(K^2) & =& 4\pi^2 \sum_i Q^2_i\left[\Pi_i(K^2)
	-\Pi_i(0)\right]\ .
\end{eqnarray}
Here $Q_i$ is the electric charge in units of $e$, and $\Pi_i(K^2)$
the vacuum polarization for the $i^{th}$ quark flavor with space-like
momentum-squared $K^2$. $\Pi(K^2)$ is logarithmically divergent and is renormalized by subtracting its value at $K^2=0$.
Since $f(K^2)$ diverges as $K^2\to 0$, we see immediately that the largest contribution to the muon \gtwo\ will come from the low momentum region; in particular the dominant contribution to the integral
comes from momenta near the muon mass, $K^2\approx m_{\mu}^2$.

The vacuum polarization for the $i^{th}$ quark flavor in the continuum is given by (no sum on $i$)
\begin{equation}
	\Pi_{\mu\nu}^i(q) = \int d^4x {\rm e}^{iq\cdot (x-y)}
	\langle J_\mu^i(x) J_\nu^i(y)\rangle\ ,
\end{equation}
where $J_\mu^i(x)=\bar\psi_i(x)\gamma_\mu\psi_i(x)$ is the
vector current.

For staggered quarks, the conserved current has the following form, in terms of the one-component staggered fields $\chi$,
\begin{equation}
	J_{\mu}^i(x) = \frac{1}{2}\eta_\mu (x)
	\left[\bar \chi_i(x+\hat\mu)U^\dagger_\mu(x)\chi(x)+
	\bar \chi_i(x)U_\mu(x)\chi(x+\hat\mu)
	\right]\ ,
\end{equation}
where $\eta_\mu (x) = (-1)^{\sum_{\nu < \mu}x_\nu}$ is the staggered fermion phase, and satisfies a discrete conservation law 
\begin{eqnarray}
\Delta_\mu J_{\mu,i} =0 
\end{eqnarray}
(up to possible contact terms depending on the correlation function) where $\Delta_\mu$ is a backward difference operator. 
This implies an exact lattice Ward-Takahashi identity for the two-point function, given by \eq{WI} with $q_\mu \to \hat q_\mu = 2 \sin (aq_\mu/2)/a$. It is important to note that this Ward-Takahashi identity must be satisfied (in a numerical simulation) on each gauge configuration separately (to numerical precision). This is true for all of the calculations presented here and serves as a strong check of the calculation.

In evaluating the quark propagators using the current above, we do not include the Naik term \cite{Naik:1986bn} appearing
in the improved staggered (asqtad) action \cite{Orginos:1998ue,Orginos:1999cr,Lepage:1998vj}
used by the MILC collaboration to generate the gauge field configurations used in this study. 
This is because the form of the Ward-Takahashi
identity in the presence of the Naik term is not as simple as it is for the nearest neighbor point-split current; an extra term appears. The
effects of not including the Naik term were discussed in
 \cite{g-2_2}.
Recall that the Naik term is meant to improve the dispersion relation of the naive staggered fermion which breaks down for large values of the momentum. Thus this omission  should not significantly effect the long-distance, low momentum physics with which we are mainly concerned. In  \cite{Bernard:1997mz} it was shown that the effect on
hadron masses, and the
vector and pion meson masses in particular, from the Naik term was negligible compared to the
statistical error for the types of parameter values used in this study, {\it i.e.}, small quark mass and lattice spacing.
While we know of no significant effect due to this omission,
it is possible that an unknown error is introduced by the 
``mixed action" we are using. Since  \cite{g-2_2} showed little difference (within statistical errors) whether or not the Naik term is present, we assume for now that any error introduced by this mixed action is negligible.

\begin{figure}[ht]
\begin{center}
\includegraphics{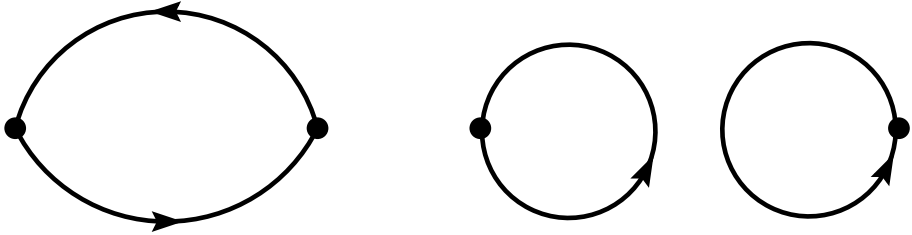}
\caption{The two Wick contractions for the vacuum polarization. The dots are insertions of the vector current $J_\mu(x)$. Only the first has been calculated in this study.}
\label{fig:contractions}
\end{center}
\end{figure}

Further details of the lattice calculation can be found in
 \cite{g-2_1,g-2_2,g-2_3,QCDSF}, although we make one last comment
here as it pertains to our discussion later. There are two types of
Wick contractions (shown in Figure~\ref{fig:contractions}) that need to be evaluated for a
complete calculation; however the disconnected diagrams shown in Figure~\ref{fig:contractions} have not been
computed due to the well-known practical difficulty in calculating them. The disconnected diagrams, when summed over $u$, $d$, and $s$ quarks, cancel in the SU(3) flavor limit, and they are suppressed according to
Zweig's rule. As such we do not expect them to make a significant
contribution to the final result, so for now we leave the calculation of the disconnected diagrams for a future work.

Table~\ref{tab:latt_params} shows the lattice parameters used in this
calculation. We employ the ``fine'' MILC configurations ($a=0.086$ fm) generated with the asqtad fermion action with
the light valence quark mass set equal to the degenerate light sea quark mass and fixed value for the
strange valence quark mass equal to the strange sea quark mass\footnote{This is nominally the ``physical'' value
of the strange quark mass, although it was noted in
 \cite{MILC_SPECTRUM} that this value is slightly too high on
these lattices, and intstead $am_s\approx 0.027$ corresponds to the physical strange quark mass. This difference should, as in
Refs. \cite{MILC_SPECTRUM,MILC_CHIRAL_FITS}, have a negligible
effect.}. We will discuss the lattice data and the fits in Sec.~\ref{sec:fits}.

\section{Staggered Chiral Perturbation Theory with Photons}
\label{sec:schpt_photons}

Since the low-energy, small quark mass regime is the most important part of the lattice
calculation (because of the kernel in \eq{kernel}), we will need a precise,
theoretically sound, fitting function to model the lattice data. In principle, if the lattice volume were large enough to have arbitrarily small momenta, and the quark masses took on their physical values,
then any simple, smooth parameterization of $\Pi(K^2)$ would do,
provided the statistical errors on the lattice data were small enough.
Even on the state-of-the-art lattices used in this study, this is not the case, so the lattice calculation must be leveraged as much as possible with a good fit function.
Such a function for the vacuum polarization
can be determined in chiral perturbation theory (\chpt) coupled to photons treated as
external fields, as in  \cite{GASSER_LEUTWYLER}, for example. Coupling \chpt\ to photons is simple: to treat the photons as
external fields we merely replace the ordinary derivative with a
covariant one. The generalization to staggered \chpt\ (\schpt)
is also relatively straightforward, but 
there may be additional lattice artifacts (``taste-symmetry\footnote{Each staggered fermion corresponds to four degenerate Dirac fermions in the continuum limit. These are called tastes, in analogy to flavors.} violations") that arise from the interactions of the pions with the photons. As we shall see, these terms do not arise at the order to which
we work.

First let us write down the staggered chiral Lagrangian in the
presence of photons. The chiral symmetry for three flavors of
staggered quarks is an $SU(12)_L\timestwo SU(12)_R$
symmetry. We incorporate the pseudo-Goldstone bosons in the field
$\Sigma$, where $\Sigma\to L\Sigma R^\dagger$ under a chiral
transformation, with $L\in SU(12)_L$ and $R\in SU(12)_R$.

In Minkowski space, we have the Lagrangian
\begin{equation}\label{eq:PionLag}
  \cL_2  = 
  \frac{f^2}{8}\Tr[D_\mu\Sigma (D^\mu\Sigma)^\dagger]
  + \frac{\mu f^2}{4}\Tr[\cM\Sigma + \Sigma^\dagger\cM] 
  - a^2\cV\, ,
\end{equation}
where $\cV$ is the taste-symmetry breaking potential for multiple
flavors of staggered quarks \cite{SCHPT}, $\cM = {\rm diag}(m_u
1_{4\timestwo 4}, m_d 1_{4\timestwo 4}, m_s 1_{4\timestwo 4})$ is the
light quark mass matrix, $f$ is the tree-level pion decay constant
(normalized here so that the physical $f_\pi\approx 131$ MeV), and $\mu$ is a
constant with dimensions of mass. The subscript on $\cL_2$ signifies the ${\cal O}(p^2,m_\pi^2, a^2)$ chiral Lagrangian\footnote{Generally, for staggered fermions the leading discretization errors begin at ${\cal O}(a^2)$.}. The explicit form of $\cV$ is given
in  \cite{SCHPT}, and for our purposes gives rise to the
splitting of the various taste mesons. It additionally gives rise to
two-point interaction terms which mix the flavor-neutral vector- and
axial-taste mesons, but these will not have any effect on the present 
calculation.  The covariant derivative is defined as 
\begin{eqnarray}
D_\mu\Sigma = \partial_\mu\Sigma + i e A_\mu [Q,\Sigma],
\label{eq:covar deriv}
\end{eqnarray}
where $Q$ is the light
quark charge matrix in units of $e$, and $A_\mu$ is the photon field.

Note that we have dropped the term in the Lagrangian which is
proportional to $m_0^2$. This term, due to the anomaly,
gives the taste- and flavor-singlet meson 
 ($\eta'_I$) a large mass which we send to infinity, as generally the
effects of the $\eta'_I$ are decoupled from the low energy
regime. In addition, the $\eta'_I$, along with all the other flavor-neutral pions,
does not contribute to the photon vacuum polarization.

The question now is whether or not there will be additional terms 
which can violate the taste symmetry at the order to which we work,
but which include photon interactions. 
Due to the gauge structure of the theory, the only place where
the gauge fields can enter is either through a derivative term,
or by inserting terms that contain the field strength tensor $F_{\mu\nu}$ in a Lorentz- and gauge- invariant way. 
The first thing to note is
that $\cV$ does not include derivative interactions, so there
are no photon terms arising from covariant derivatives in $\cV$.  Terms with only a single factor of $F_{\mu\nu}$ are not allowed due to Lorentz invariance. Thus, in the $O(a^2)$ Lagrangian, there are no taste-violating terms that contain photons.

As usual in chiral perturbation theory, working at one-loop order in $\cL_2$ requires us to include tree-level terms from $\cL_4$, as these two sets of terms are of the same order in power counting. Additionally, these tree-level terms absorb the infinities (and scale dependence) that arise from the one-loop integrals.
The only terms from $\cL_4$ that violate the taste symmetry are ${\cal O}(a^2 m_\pi^2,a^2p^2, a^4)$ since we must include factors of $a^2$. Terms in the continuum $\cO(p^4)$ Lagrangian \cite{GASSER_LEUTWYLER} also contribute at this order.

In \cite{SCHPT_NNLO} all taste-symmetry breaking operators at ${\cal O}(a^2 m_\pi^2,a^2p^2, a^4)$ which include external fields along with the meson field have been enumerated, and thus should include all possible terms that can contribute in the present case. 
It is easy to see that none of the terms listed in
 \cite{SCHPT_NNLO} contributes here, just as many of the
${\cal O}(p^4)$ terms do not. The reason is that any terms which include a covariant derivative on $\Sigma$ vanish when $\Sigma=1$ in the ${\cal O}(a^2 m_\pi^2, a^2p^2, a^4)$ terms\footnote{Recall that we are calculating the photon two-point function, so to obtain a tree-level term which will contribute, then we can only have factors of the photon field. Thus, we will ultimately set $\Sigma=1$ in all $\cL_4$ terms.} (there
can not be any factors of momenta if each term including a field strength tensor is to have the correct dimensions). 
Thus, all operators listed in  \cite{SCHPT_NNLO} vanish, and from the ${\cal O}(p^4)$ terms \cite{GASSER_LEUTWYLER}, 
we only have the following terms which will contribute
\begin{equation}\label{eq:p4_terms}
  \cL_4  =  L_{10} \Tr[\Sigma^\dagger F_{\mu\nu}\Sigma F^{\mu\nu}]
  + 2H_1\Tr[F_{\mu\nu}F^{\mu\nu}] \ .
\end{equation}
We can easily see that both of these terms will contribute in the same
fashion when we set $\Sigma=1$. 

Terms with only external fields (contact terms) were not considered in
 \cite{SCHPT_NNLO}, although in the current calculation they could
play a role, just as the $H_1$ term in \eq{p4_terms} does. We can see that explicit taste-symmetry breaking terms of this form cannot arise. Any terms
included here have to be ${\cal O}(a^4,a^2p^2)$, but two factors of $F_{\mu\nu}$
mean any such contact term begins at ${\cal O}(p^4)$\footnote{Note the external photon field
$A_\mu$ is ${\cal O}(p)$ in the chiral power counting scheme 
\cite{GASSER_LEUTWYLER}, so $F_{\mu\nu}\sim \partial_\mu A_\nu - \partial_\nu A_\mu$ is ${\cal O}(p^2)$.}. Thus,
at most one power of
$F_{\mu\nu}$ can appear in each term. Lorentz-invariant terms with one factor of $F_{\mu\nu}$ are not allowed, as
they require more derivatives. However, it could be possible to break the Lorentz symmetry by coupling the Lorentz indices on the field strength tensor to the vector indices on the taste matrices, as in
Refs.~\cite{SCHPT_NNLO,SCHPT_HL}. Following a similar spurion analysis, we find that the only terms that would arise are those which involve the diagonal elements of the field strength tensor, which all vanish.
Thus there are no ${\cal O}(a^4,a^2p^2)$ taste violating terms. We could get ${\cal O}(a^2)$ corrections to various parameters (such as $L_{10}$ or $H_1$), but these would not arise from violations of the taste symmetry, but instead from other lattice discretization errors.

While the above argument is purely technical in nature (enumerating all terms that are allowed to appear in the theory), we can see from a more physically intuitive picture that there are no taste violations (at least to this order) from photon interactions. Taste violations generally arise from gluonic exchange with momentum near the cutoff. These could also arise from photon exchange with momentum near the cutoff, but would be suppressed compared to the gluon exchange by a factor $\alpha/\alpha_{\rm s}$. More importantly, however, in this lattice calculation these interactions do not occur because we do not treat the photons as dynamical fields. In other words, our simulation is a quenched QED simulation, and as such we cannot have such taste violations arise. This does not mean there are not any taste violations here; they can, and do, arise in the meson mass splittings which we already know come only from high momentum gluon exchange.

\subsection{One-loop calculation with pions}

Using \eq{PionLag}, it is straightforward to calculate the one-loop contribution to the
photon vacuum polarization. As usual we write $\Sigma =
\exp\left[i\Phi/f\right]$, where $\Phi$ is the $12\timestwo 12$ matrix
of pseudo-Goldstone bosons
\begin{eqnarray}\label{eq:Phi}
	\Phi & = & \left( \begin{matrix}
	U & \pi^+ & K^+ \\
	\pi^- & D & K^0 \\
	K^- & \bar K^0 & S
	\end{matrix}\right)\ ,
\end{eqnarray}
where $\Phi$ is written in the flavor basis so $U$ is a $\bar u u$ meson,
$D$ is a $\bar d d$ meson, and $S$ is an $\bar s s$ meson. Also, each
element shown in \eq{Phi} is a $4\timestwo 4$ matrix, written as, for
example, $U=\sum_a U_a T^a$, where 
\begin{equation}\label{eq:taste_mats}
	T^a \in \{ \xi_5, i\xi_{\mu5}, i\xi_{\mu\nu}, \xi_{\mu}, 
	1_{4\timestwo4} \}\ 
\end{equation}
is a Hermitian taste
matrix ($a=1\ldots 16$),
$1_{4\timestwo4}$ is the $4\timestwo 4$ identity, the $\xi_\mu$ are Hermitian Dirac matrices which satisfy $\{\xi_\mu,
\xi_\nu\} = 2\delta_{\mu\nu}$, $\xi_{\mu\nu}$ is defined as
$\xi_\mu\xi_\nu$ (similarly with $\xi_{\mu 5}$), and $\mu<\nu$ in
\eq{taste_mats}. Note that although we are in Minkowski spacetime, we use Euclidean taste matrices as in  \cite{SCHPT_HL}.

\begin{figure}[ht]
\begin{center}
\includegraphics{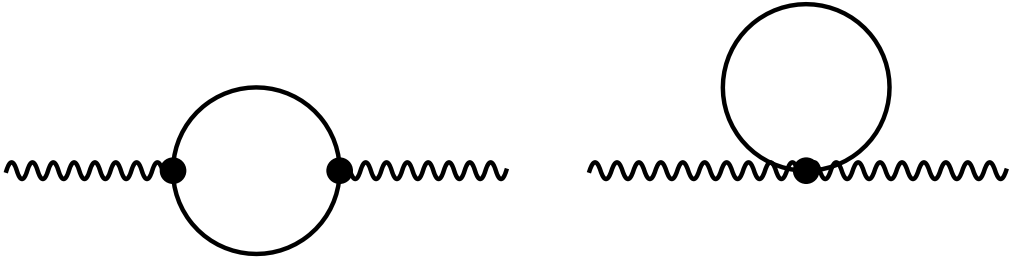}
\caption{The two one-loop diagrams contributing to the photon vacuum polarization in \schpt. The solid line can be either a charged pion or a charged kaon.}
\label{fig:one_loop_pion}
\end{center}
\end{figure}

Expanding $\Sigma$ to quadratic order in $\Phi$ gives two types of
couplings to the photon, as is the usual case for scalar QCD. Both vertices have two pions, and one vertex has a single photon while the other has two photons.  From these vertices, we see
that there are two possible diagrams, shown in Figure~\ref{fig:one_loop_pion}, where the mesons
in the loops are only the charged pion or kaon, of any of the 
16 tastes. 
Note that although the
second diagram is a constant in $q^2$, where $q$ is the external
momentum, it is required for gauge invariance of the total amplitude
at this order.

The calculation for these two diagrams is rather straightforward, and
we get the following result, after continuing the external momentum to
Euclidean space, $q^2\to -q^2$,
\begin{equation}\label{eq:Pi_pion_4tastes}
	\Pi^{{\rm 1-loop}}_{4, \pi}(q^2)  =  \frac{\alpha}{4\pi}
	\sum_{M, t} \left[
   	 G\left(\frac{4m^2_{M_t}}{q^2}\right) 
       - \frac{8}{9} + 
     \frac{1}{3}\ln\left(\frac{m_{M_t}^2}{\Lambda^2}\right)
      \right]\ ,
\end{equation}
where 
\begin{equation}
	G(x) = \frac{1}{3}\left(1+ 
   	      x\right)^{3/2}
   	    \ln\left( 	  
	    \frac{\sqrt{1+x}+1}{\sqrt{1+x}-1}
       \right)-\frac{2x}{3}\ ,
\end{equation}
$M\in\{\pi^+,K^+\}$, and the sum over $t$ is over the 16 tastes of
mesons. $\Lambda$ is the chiral cutoff scale. The subscript $4$ refers
to the fact that we have not yet taken into account the reduction from
$4\to1$ tastes per flavor which must be done to arrive at the physical
number of mesons. The mass for a meson $M_t$ consisting of two
quarks $a$ and $b$ and with taste $t$ in terms of the Lagrangian
parameters is
\begin{equation}\label{eq:pion-masses}
	m^2_{M_t} = \mu (m_a + m_b) + a^2 \Delta_t
\end{equation}
where $\Delta_t$ is a combination of various parameters from $\cV$,
and can be found in  \cite{SCHPT}. The measured values on the
lattice can be found in  \cite{MILC_SPECTRUM}.

The above expression is the lowest order result for the vacuum polarization in chiral perturbation theory. In particular, to this order, it is a valid description of QCD and needs no alterations whether the disconnected diagrams are included in the lattice calculation or not. 
Imagine a lattice calculation in the $SU(3)$ flavor limit; the disconnected diagrams in the electromagnetic current correlation function cancel exactly. However, in Eq.~\ref{eq:Pi_pion_4tastes}
the only change is to set $m_\pi=m_K$. In other words, effects from these diagrams
must vanish like $(m_\ell-m_s)^2$ and hence are higher order. 
	
So far, \eq{Pi_pion_4tastes} includes the effects of each flavor of
quark having four tastes. In the simulations, the fourth root of
each quark determinant is (formally) taken to remove the extra
unphysical tastes, so we need to include these effects by hand in \schpt\ \cite{SCHPT, CHIRAL_FSB}. To do this, we use
standard quark flow techniques (for example,  \cite{SCHPT} and
references therein), with minor adaptations due to the presence of the external photon field.

\begin{figure}[ht]
  \centering
  \includegraphics[width=3in]{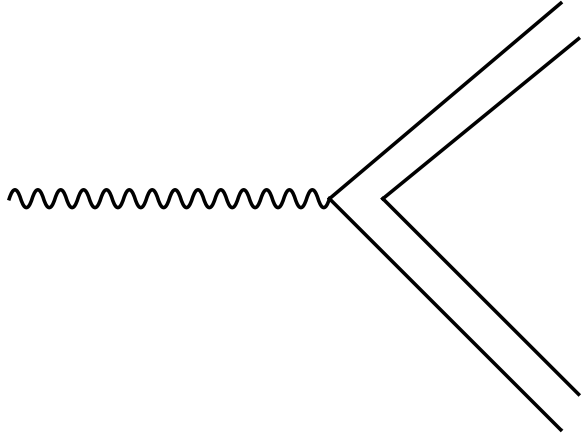}
  \caption{Quark-level vertex coming from the 
  one-photon term in the Lagrangian. The solid lines are quarks.}
  \label{fig:em_vert1}
\end{figure}

\begin{figure}[ht]
  \centering
  \includegraphics[width=3in]{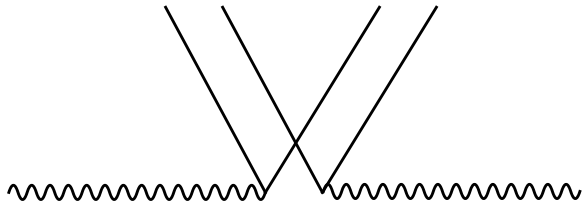}
  \caption{Quark-level vertex coming 
  from the two-photon term in the Lagrangian. 
  As in Figure~\ref{fig:em_vert1}, the solid lines are quarks.}
  \label{fig:em_vert2}
\end{figure}

As mentioned before, the interaction terms involve two pion legs and either one or two photon legs. The quark-level diagrams for these two vertices are shown in Figures~\ref{fig:em_vert1} and \ref{fig:em_vert2}. Combining these vertices in such a way to correspond to the diagrams in Figure~\ref{fig:one_loop_pion}, we see that both diagrams include \textit{two internal quark loops}. Thus, every term in our result must be multiplied by a factor of $(1/4)^2 = 1/16$ to account 
for the transition from $4\to1$ tastes per flavor.\footnote{Note that for the vacuum polarization computed in the chiral theory 
	all quarks involved are
	``sea'' quarks since we compute the photon two-point function, 
	even though in the lattice calculation there 
	is a distinction between sea and valence quarks. Thus, 
	although one might assume that some quark lines are treated 
	as valence quarks and do not get a factor of 1/4, this is 
	in fact not the case. This is also the reason why, although 
	we exclude disconnected diagrams in the lattice calculation, 
	we include what look like, at the quark-flow level, disconnected 
	diagrams. Both of these statements can be made concrete by 
	treating the \chpt\ calculation in a partially quenched 
	setting, and one finds the same results presented here.} 
Incorporating the necessary factors of $1/4$ in our expression above, and correcting for the exclusion of the disconnected diagrams, we have
\begin{equation}\label{eq:Pi_pion_1taste}
	\Pi^{\rm 1-loop}_{1,\pi}(q^2)  =  \frac{\alpha}{4\pi}
	\left\{
      \frac{1}{16}\sum_{M, t} \left[
   	 G\left(\frac{4m^2_{M_t}}{q^2}\right) 
       - \frac{8}{9} +
     \frac{1}{3} \ln\left(\frac{m_{M_t}^2}{\Lambda^2}\right)
       \right] +A\right\}\ .
\end{equation}
The constant $A$ includes the analytic terms from the ${\cal O}(p^4)$
Lagrangian (shown in \eq{p4_terms}) as well as terms that can
arise from non-taste-violating ${\cal O}(a^2p^2)$ operators mentioned above. Since the current work is only done at a single lattice spacing, we do not worry about the explicit $a^2$ dependence in $A$, although calculations at multiple lattice spacings would require writing
\begin{equation}
	A = A_0 + A_1 a^2\ ,
\end{equation}
with $A_0 \propto L_{10} + 2H_1$, and $A_1$ some coefficient parametrizing the discretization errors. Note that this additional discretization error does not affect the muon \gtwo, as it will be subtracted off when renormalizing the final result. The only relevant discretization errors come from the different meson masses in \eq{Pi_pion_1taste}.

We would like to examine \eq{Pi_pion_1taste} in both the zero momentum
and the chiral limits. These correspond respectively to $x\to\infty$
and $x\to 0$ in the function $G(x)$. The first limit is well defined,
and we have that
\begin{equation}
	\lim_{x\to\infty}G(x) = \frac{8}{9}\ .
\end{equation}
However, in the chiral limit, when $x\to 0$, the function $G(x)$ diverges like
\begin{equation}
	\lim_{x\to 0}G(x) = \frac{1}{3}\ln\left(\frac{4}{x}\right) 
	+ O(x) \ .
\end{equation}
This divergence is exactly canceled in the full vacuum polarization by the term $\frac{1}{3}\ln(m^2_{M_t}/\Lambda^2)$, so 
$\Pi^{\rm 1-loop}_{1,\pi}(q^2)$ is finite in the chiral limit, as we expect. 

\begin{figure}[ht]
\begin{center}
\includegraphics[width=6in]{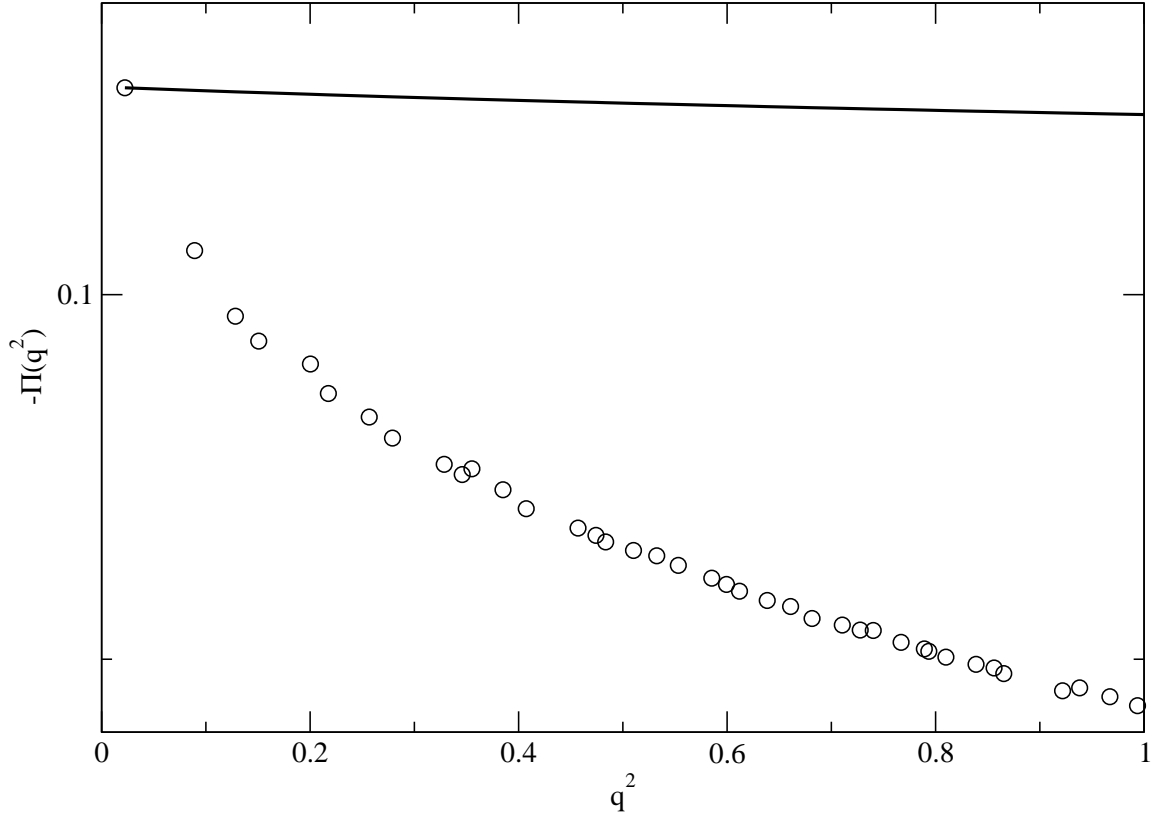}
\caption{Comparison of \eq{Pi_pion_1taste} to the lattice vacuum polarization, where we have matched the two at the smallest value of $\hat q^2$ to fix the overall constant. The open circles are the datapoints (errors not shown) for the lightest quark mass used, $am_l=0.0031$. The solid line is the \schpt\ curve.}
\label{fig:onlypion}
\end{center}
\end{figure}

We find immediately that this expression does not
describe the data at all. We can fix the overall constant $A$ by matching
the data at the lowest value of the lattice momentum, $q^2=(2\pi/aT)^2\approx 0.0223$ GeV$^2$, for the largest lattice used in our
study and
use the values of the different meson masses from
 \cite{MILC_SPECTRUM}. After fixing the overall constant there are no free parameters, and the resulting fit function and the lattice
data for light quark mass $am_l=0.0031$ are shown in Figure~\ref{fig:onlypion}. The slope for low $q^2$ is smaller by roughly two orders of magnitude in the fit function compared to the data. It is obvious that this cannot be a complete description of the
low-energy physics for the photon vacuum polarization.

We can use our knowledge of the quenched simulations \cite{g-2_1,QCDSF} to understand this. In the quenched case, the entire result for the vacuum polarization is described by a form dictated by vector particle intermediate states since they are absolutely stable in this case. We imagine that the vectors can also make a significant contribution to the photon vacuum polarization in the dynamical case, so we should include them in our calculation. Thus, the next section discusses how we incorporate vectors in \schpt.

\section{Including vectors in \schpt\ as resonances}\label{sec:vector_schpt}

Including the lightest vector mesons such as the $\rho$ or the
$\omega$ is more tricky than including the light
pseudoscalars. Unlike the pions, the light vectors have no special role when discussing chiral symmetry. This is not an inherent problem, although there could be some model dependence which arises depending on how one includes the effects of the vectors. Here we use the formalism of
Ecker, \et\ in  \cite{Ecker:1988te}, treating the vectors
as resonances in the effective field theory. This describes the vectors with antisymmetric Lorentz tensor fields, and not with Lorentz vector fields. With the correct free-field Lagrangian, this will enforce the correct number of degrees of freedom, and is useful when coupling the vectors to the photon \cite{Ecker:1988te}. The particular form of the vector field should not alter the physics, as was discussed in Ref.~\cite{Ecker:1989yg}.
\footnote{By including the vector resonances, we emphasize that we are now resorting to a model, so in principle there can be model dependence in the results. However, at present this model is required to adequately fit the low momentum region of the lattice data and as such need only furnish a smooth function that accurately describes that data. When larger volumes are used, one might dispense with such fits as the results could be reliably extracted from the actual data if enough low-momentum points are reached.}

The octet of light vector mesons
is an antisymmetric spin-1 field, $V_{\mu\nu}$, which is an
$SU(12)$ matrix. The continuum free-field Lagrangian in Minkowski space is
\begin{equation}\label{eq:rho_free_lag}
	\cL_v = -\frac{1}{2}\Tr\left[\partial^\mu V_{\mu\nu}
	\partial_\rho V^{\rho\nu}
	- \frac{1}{2}m_V^2 V_{\mu\nu}V^{\mu\nu}\right]\ ,
\end{equation}
where $m_V$ is the vector meson mass.

From the point of view of quark content, the vector fields have the
same status as the light pseudoscalar fields. For example, a $\rho^+$
has the same light quark content as a $\pi^+$, although the relative
spins of the constituent quarks are different. There is a fundamental
difference from the point of view of chiral symmetry, in that the
vectors do not have any special properties when the light quark masses
vanish. What is important, however, is that the vectors and the pions
have the same transformation property under the unbroken vector
$SU(12)$ subgroup, as this is governed solely by the quark content. The primary difference is that we cannot determine the leading dependence of the vector mass $m_V$ on the quark masses from chiral symmetry alone, as we can for the pions.

As is standard in
\chpt, we expect the $SU(12)_L\timestwo SU(12)_R$ chiral symmetry to
spontaneously break down to the vector subgroup $SU(12)_V$. It is
under this subgroup that the vector and pseudoscalar fields will have
the same transformation property
\begin{eqnarray}\label{eq:vec_subgroup_trans}
	\Sigma &\to & U_{V}\Sigma U_{ V}^\dagger\ , \nonumber\\
	V_{\mu\nu} & \to & U_{V} V_{\mu\nu} U_{V}^\dagger\ ,
\end{eqnarray}
where $U_{V}\in SU(12)_V$ is a constant vector transformation. With regard to the underlying chiral symmetry, any
transformation which reduces to \eq{vec_subgroup_trans} in the
appropriate limit will lead to the same physics. 
A simple choice
for the underlying transformations is made by defining $\sigma^2 = \Sigma$,
where under $SU(12)_L\timestwo SU(12)_R$ we have
\begin{equation}\label{eq:sig_trans}
	\sigma \to  L \sigma U(x)^\dagger 
	= U(x)\sigma R^\dagger\ .
\end{equation} 
$U(x)$ is an element of $SU(12)$, and since it depends on the pion fields, we see that
it is position-dependent (the explicit form of $U(x)$ will not be needed). 
We can now define a transformation rule for the vector octet as
\begin{equation}\label{eq:rho_trans}
	V_{\mu\nu} \to U(x) V_{\mu\nu} U(x)^\dagger \ .
\end{equation}
When $SU(12)_L\timestwo SU(12)_R\to SU(12)_V$, we set $U(x)=L=R=U_{V}$,
and all the fields have the correct transformations under the unbroken
vector subgroup. 

The interaction terms to lowest order are given by
\begin{equation}\label{eq:rho_int}
  \cL_\rho = \frac{f_V}{2\sqrt{2}}
  \Tr\left(V_{\mu\nu}f_+^{\mu\nu}\right)  + \frac{iG_V}{2\sqrt{2}}
  \Tr\bigl(V_{\mu\nu}[\sigma^\mu, \sigma^\nu]\bigr)  \ ,
\end{equation}
where $f_V$ and $G_V$ are real couplings ($f_V$ is the tree-level
vector decay constant). We use the definitions
\begin{eqnarray}
  \sigma^\mu & = & i\sigma^\dagger D^\mu\Sigma \sigma^\dagger\ ,\\
  f^{\mu\nu}_+ & = & \sigma F^{\mu\nu}\sigma^\dagger
  + \sigma^\dagger F^{\mu\nu}\sigma\ , \\
  F^{\mu\nu} & = & eQ(\partial^\mu A^\nu - \partial^\nu A^\mu)\ .
\end{eqnarray}
The second term in \eq{rho_int} will not affect the present
calculation. An explicit form for the vector field is given by
\begin{eqnarray}\label{eq:rho_field}
  V_{\mu\nu} & = & \left(
  \begin{matrix}
    \frac{1}{\sqrt{2}}\rho^0 + \frac{1}{\sqrt{6}}\omega_8
    & \rho^+ & K^{*+}\\
    \rho^- & -\frac{1}{\sqrt{2}}\rho^0 + \frac{1}{\sqrt{6}}\omega_8
    & K^{* 0}\\
    K^{* -} & \overline{K}^{*0} &
    -\frac{2}{\sqrt{6}}\omega_8
  \end{matrix}\right)_{\mu\nu}\ ,
\end{eqnarray}
where in this expression, each entry is a $4\timestwo 4$ matrix,
corresponding to the 16 tastes of each vector meson.

We pause our discussion for a comment about the different tastes of a
staggered vector meson. There are sixteen tastes, falling into eleven
multiplets which are irreducible representations of the lattice
symmetry group \cite{Golt_stag}. Each of these irreps, in principle, will have
different masses, but empirically the mass differences of the tastes
are negligible. 
We use this fact to ignore these mass differences,
to a first approximation, in the current analysis. Of course this does
not imply that there can be no significant effects from taste-changing
interactions among the vector particles in general, although we assume
they give rise to errors which are smaller than the
statistical and systematic errors already present in our calculation.

Finally, we should include terms in the vector part of the Lagrangian that could violate the taste symmetry, but we choose to neglect them for the following reasons. On one hand, taste violations arise through additional terms in the masses of the vectors (as happens with the pions), and these we neglect because empirically they are negligible. The other source of taste violations is additional terms that do not contribute to the tree-level masses but arise in one-loop calculations just as the flavor-neutral hairpins arise in pion masses and decay constants \cite{SCHPT,FPI}. We expect these contributions to be similar in size to the mass splittings (as in the pion sector), and thus negligible.

\subsection{Tree-level vector contribution}

\begin{figure}[ht]
\begin{center}
\includegraphics{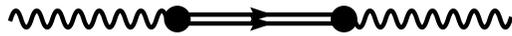}
\caption{Tree-level diagram for a vector meson which contributes to the vacuum polarization. The double-line can be either the $\rho^0$ or the $\omega_8$.}
\label{fig:rho_tree_level}
\end{center}
\end{figure}

The tree-level calculation of the vacuum polarization involving the
vectors is straightforward. We set $\sigma=1$ in the above expressions
and find the vertices which couple the vectors to the photon. There are
two two-point vertices: one where the photon couples to the
$\rho^0$ and another where the photon couples to the $\omega_8$. 
The relevant diagram is shown in Figure~\ref{fig:rho_tree_level}, and from it, we get the
following contribution to the photon vacuum polarization, after
continuing the external momentum to Euclidean space,
\begin{equation}\label{eq:rho_tree_level}
	\Pi^{\rm tree}_V(p^2) =
	-\frac{\alpha}{4\pi}\frac{(4\pi f_V)^2}{3}\left[
	\frac{3}{p^2+m^2_V} + \frac{1}{p^2+m^2_V}\right]
	=
	-\frac{\alpha}{4\pi}\frac{4}{3}
	\frac{(4\pi f_V)^2}{p^2+m^2_V}
	\ .
\end{equation}
In this expression, we have not included an explicit taste index on
the vector mesons since we are neglecting the splittings between the
various tastes. The full analysis requires making the transition from four to one tastes per flavor. By examining the Lagrangian above a similar analysis to the pion case leads to an overall factor of $1/16$ times a sum over the 16 tastes of vector mesons. Neglecting taste violations in the vector masses, the 16 degenerate contributions cancel the factor of $1/16$.

If we were to measure the decay constant as well as
the vector masses on the lattice, then this expression has no free
parameters. Notice that at this (lowest) order, the masses of both the $\rho^0$ and $\omega_8$ are the same, $m_V$. We can set this equal to the $\rho$ mass, and any discrepancies enter at ${\cal O}(p^6)$ \cite{Ecker:1988te}.

A quick analysis shows that for small momentum, the tree-level
$\rho$ contribution dominates the one-loop pion contribution
above. Expanding \eq{rho_tree_level} around $p^2=0$, we get
\begin{equation}\label{eq:rho_tl_exp}
	\Pi^{\rm tree}_V(p^2) =
	\Pi^{\rm tree}_V(0)
	+ \frac{\alpha}{4\pi}\frac{4(4\pi f_V)^2}{3}
	\frac{p^2}{m^4_{\rho^0}}+\cO(p^4)
	\ ,
\end{equation}
while for the one-loop pion contribution we get
\begin{equation}\label{eq:Pi_pion_1taste_expanded}
	\Pi^{\rm 1-loop}_{1,\pi}(p^2)  =
	\Pi^{\rm 1-loop}_{1,\pi}(0)
	+  \frac{\alpha}{4\pi}
      \frac{1}{16}\sum_{M, t} 
      \frac{p^2}{30m^2_{M_t}}
     + \cO(p^4) \ .
\end{equation}
To more easily compare these two expressions, let us first subtract
off the zero momentum contribution, because it is the slope of the
low-momentum region that we are interested in. Taking the ratio of these gives
\begin{equation}\label{eq:rho_pi_ratio}
	\frac{\Pi^{\rm tree}_V(p^2) - \Pi^{\rm tree}_V(0)}
	{\Pi_\pi(p^2) -\Pi_\pi(0)} =
	\frac{640(4\pi f_V)^2}
	{\displaystyle\sum_{M, t} \frac{m^4_{\rho^0}}{m^2_{M_t}}}\ .
\end{equation}
Using a rough estimate for $f_V\approx 200$MeV
\cite{Becirevic:2003pn}, we see that only for very light pion masses
will the pion contribution dominate. For the ensemble with the light quark mass $am_l = 0.0062$, we see that
\begin{equation}
	\sum_{M, t} \frac{1}{m^2_{M_t}}
	 \approx 200 {\rm GeV}^{-2},
\end{equation}
and so for a vector mass of $\approx800$MeV, \eq{rho_pi_ratio} gives  about 50. In the practical simulations we expect the vector contribution to dominate. Even at the physical masses in the continuum limit, this ratio is still roughly 10, so although the pion contribution becomes more important for lighter quark masses, the majority of the slope of the vacuum polarization comes from the vectors.

Although the contribution of \eq{Pi_pion_1taste} is much smaller that
of the $\rho$, we will include it to determine the effects it has on
the final fits. As such, we will have to include also the one-loop
diagrams which involve the $\rho$ for consistency. This is done in the
next section.

\subsection{One-loop terms with the $\rho$}

\begin{figure}[ht]
\begin{center}
\includegraphics{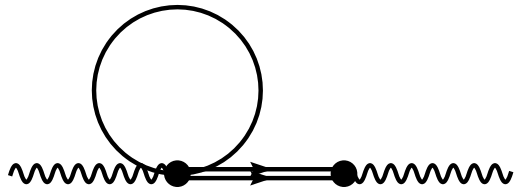}
\caption{The one-loop diagram involving the vector mesons. The solid line is either a pion or kaon, and the double solid line is the vector meson.}
\label{fig:rho_one_loop}
\end{center}
\end{figure}

The only diagram that will contribute here is 
shown in Figure~\ref{fig:rho_one_loop} (there is a second, with the pion tadpole attached to the other vertex). There is also the possibility of a term arising as a self energy correction to the vector (so the vector line would have a pion loop inside). This has been evaluated (Eq.~(28) of  \cite{Rosell:2004mn}), but it is $\cO(p^4)$. Although this term is proportional to an unknown low energy constant that in principle could be rather large, to include it consistently would require us to include higher order terms from the pure pion sector. As such, it is consistent to neglect this piece.

A straightforward calculation gives the one-loop terms. Putting
this together with the tree-level vector contribution we get the
complete one-loop result for the vectors,
\begin{eqnarray}\label{eq:one_loop_rho}
	\Pi_{\rm vec}^{\rm 1-loop}(p^2) 
	& = &
	-\frac{\alpha}{4\pi}\frac{(4\pi f_V)^2}{3}\Biggl\{
	\frac{3}{p^2+m^2_V}\left[1 
	-\frac{4 }{ (4\pi f)^2}\frac{1}{16}\sum_t
	\left[ 2\ell(m_{\pi_t}^2) + \ell(m_{K_t}^2)\right]\right]
	\nonumber\\&&{}
	+ \frac{1}{p^2+m^2_V}\left[1
	-\frac{12 }{ (4\pi f)^2}\frac{1}{16}\sum_t
	\ell(m_{K_t}^2)
	\right] + \frac{C_V}{p^2+m_V^2} (3m_\ell + m_s)\Biggr\}\ ,
\end{eqnarray}
where the usual chiral logarithm is
\begin{equation}
	\ell(m^2) = m^2
	\ln\left(\frac{m^2}{\Lambda^2}\right)\ ,
\end{equation}
with $\Lambda$
the chiral scale, as before. The factors of $1/16$ arise in the same way as in the purely pion quark-flow analysis. As we are neglecting taste violations from the $\rho$, we can treat it as a taste singlet, and then the $\rho$-$2\pi$-photon vertex is identical in quark-level terms to the 2 photon-2 pion vertex. Thus, the quark-level diagrams are identical to those in Figure~\ref{fig:em_vert2}, with one of the photon lines taken to be a $\rho$ line. We again keep $m_V = m_{\rho^0} = m_{\omega}$, as the discrepancies are ${\cal O}(p^6)$ and are still higher order than this expression \cite{Ecker:1988te}. 
We have included an unknown coefficient, $C_V$, which would arise from analytic terms from the higher-order Lagrangian. This is necessary to cancel the scale dependence coming from the chiral logarithms.

We can see that this is
roughly the same order of 
magnitude as the one-loop pion contribution, so if we were to include the one-loop pion
expression in our final result, we would in principle need to include this.  We do
not explicitly include a constant term in \eq{one_loop_rho}, even
though to this order it would be consistent to include such a
term. Following the reasoning as in the pion sector, such a term
would arise in the same manner. Since the full
one-loop result would include \eqs{Pi_pion_1taste}{one_loop_rho}, this
would amount to a change in the definition of the constant $A$ in
terms of the underlying chiral parameters. As we are not explicitly
writing down the higher-order terms in the Lagrangian in the $\rho$
sector, nor are we trying to extract the values of specific low-energy constants, we only need the mass and momentum dependence. 
Additionally, in the end this does not
matter since to determine the muon \gtwo, we subtract off the value of
the vacuum polarization at zero momentum, so any overall constant does
not affect the physics. Note also, though, that the one-loop corrections in \eq{one_loop_rho} could be absorbed into the definition of the parameter $f_V$, as there is no momentum dependence in these terms. We will discuss this further in the next section when we discuss the fits to the lattice data. 

\section{Numerical Results}\label{sec:fits}

\begin{figure}[ht]
\begin{center}
\includegraphics[width=6in]{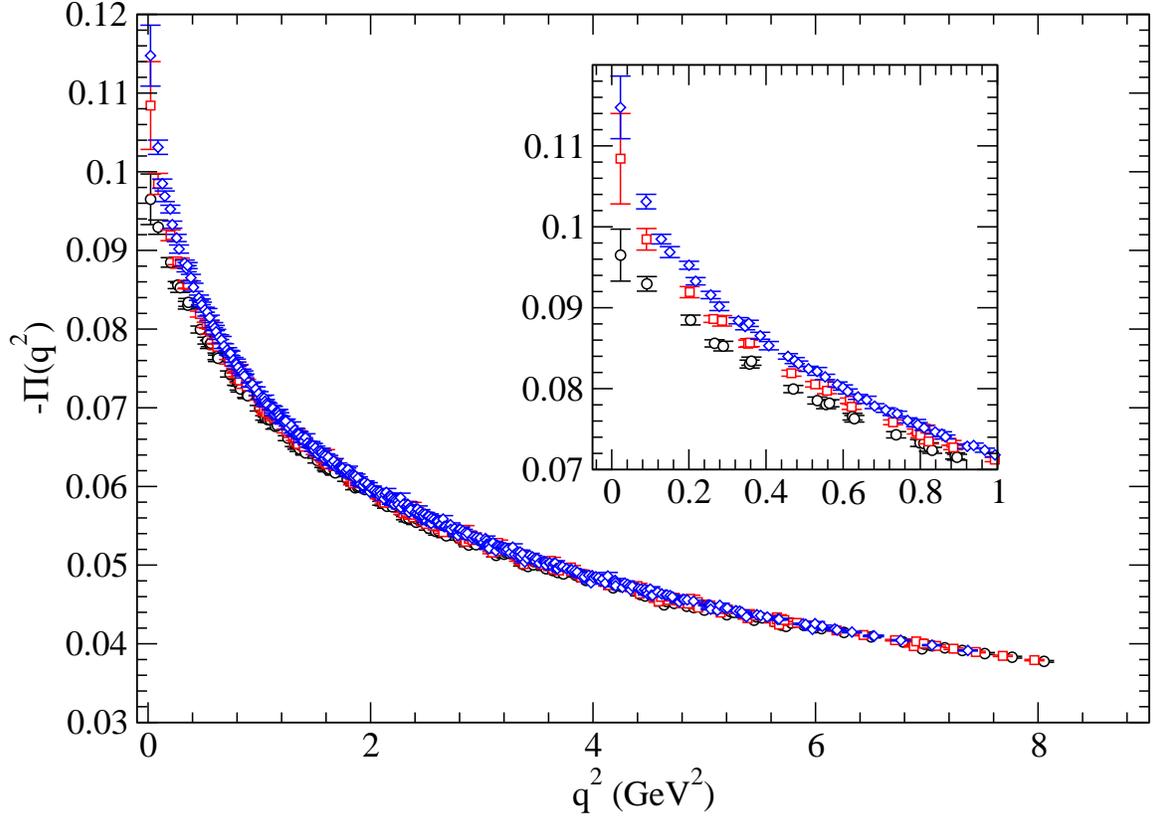}
\caption{Minus the vacuum polarization for 2+1 flavors of quarks for each light quark mass studied in this work, 0.0031 (diamonds),
0.0062 (squares), and 0.0124 (circles). The insert shows a blow up of the important low $\hat q^2$ regime. The strange quark mass is fixed to 0.031 in each case.}
\label{fig:vacpol}
\end{center}
\end{figure}

We plot the results for the 2+1 flavor vacuum polarization as a function of $\hat q^2$ in Figure~\ref{fig:vacpol}, including a close-up of the low-$\hat q^2$ region. 
As before \cite{g-2_1,g-2_2,g-2_3}, for larger $\hat q^2$, it is independent of quark mass, but as $\hat q^2\to 0$, there is a significant difference in the three plots.
$\Pi(\hat q^2)$ was computed using two quark propagator source times, 0 and 48, and results on the same lattice were averaged. We did not compute with both sources on every lattice, however, and strange and light valence quark propagators were calculated on overlapping but different subsets of each 2+1 flavor lattice ensemble (see Table~\ref{tab:latt_params}).
The statistical errors shown in Figure~\ref{fig:vacpol}, and quoted throughout, come from a single elimination jackknife procedure. Because the numbers of strange and light valence quark propagators differ within the same $am_l=0.0031$ or 0.0062 ensemble, we adopted a jackknife procedure where each light or strange quark propagator calculation of $\Pi(\hat q^2)$ is treated as a separate measurement, instead of the 2+1 flavor value of $\Pi(\hat q^2)$. This is not expected to cause difficulty since the much heavier strange quark means the light and strange quark propagators computed on each lattice are roughly uncorrelated. Also, because of the electric charges, the light quark contribution is explicitly weighted five times more than the strange quark one.
For the $am_l=0.0124$ ensemble, the jackknife error estimate for  $\Pi(\hat q^2)$ was calculated in the usual way. In all cases the errors were not significantly altered by increasing the jackknife block size to five, or by computing them using a simple binning procedure with bin size of ten configurations.

\begin{figure}[ht]
\begin{center}
\includegraphics[width=6in]{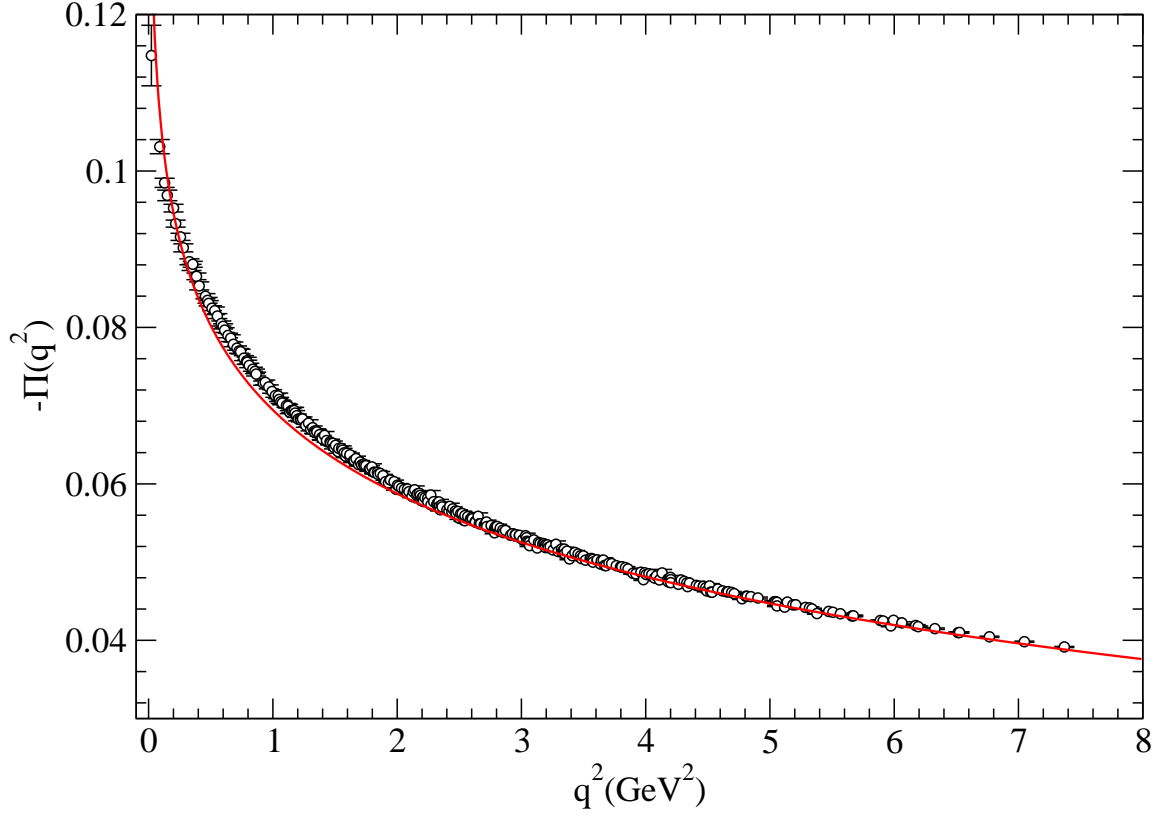}
\caption{Comparison of the 2+1 flavor vacuum polarization computed using the lattice with 3-loop continuum perturbation theory in the $\overline{MS}$ scheme \cite{Chetyrkin:1996cf}. The solid line is forced to match the lattice calculation at 2 GeV through a simple additive shift. $m_l=0.0031$ and $m_s=0.031$. The quark masses have been converted to the $\overline{MS}$ scheme using the matching factor in \cite{MILC_CHIRAL_FITS}.}
\label{fig:pert comp}
\end{center}
\end{figure}

\begin{figure}[ht]
\begin{center}
\includegraphics[width=6in]{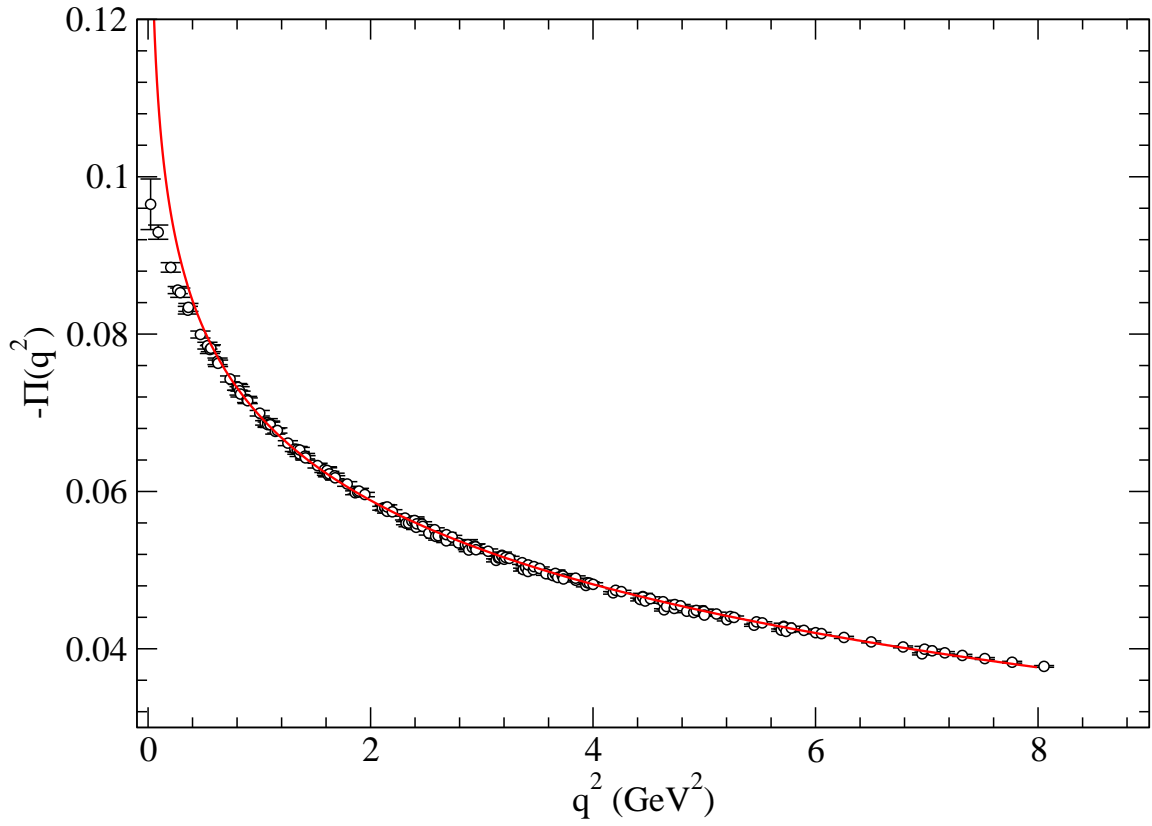}
\caption{Same as Figure~\ref{fig:pert comp} but for $m_l=0.0124$.}
\label{fig:pert comp 0.0124}
\end{center}
\end{figure}

In Figures~\ref{fig:pert comp} and~\ref{fig:pert comp 0.0124}  we compare continuum three-loop perturbation theory \cite{Chetyrkin:1996cf} with the lattice calculation of $-\Pi(q^2)$ for $0 \le q^2 \le 8$ GeV$^2$ for $m_l=0.0031$ and 0.0124, respectively. The perturbative result is given in the $\overline{MS}$ scheme, and the bare quark mass has been matched using the renormalization factor given in
 \cite{StrangeMass}. The results were forced to agree at
$\mu=2$ GeV by a imposing a simple additive shift to the perturbation theory curve. The lattice results for $3 \le q^2 \le 8$ GeV$^2$ agree impressively with perturbation theory. For lower values of
$q^2$ and $m_l=0.0031$, the lattice value increases faster until about 0.5 GeV$^2$ when
the diverging perturbation theory result overtakes it again. 
For the heavier mass, the two results coincide until about 1 GeV$^2$ where the perturbative one again overtakes the non-perturbative one.
This behavior with quark mass is mostly indicative of the variation with
quark mass in the lattice results
since the perturbation theory curve is insensitive to the light quark mass. This can also be seen from Figure~\ref{fig:vacpol}; the two lattice results are clearly separated at 1 GeV$^2$.
The above comparison is, of course, not unique. We could have matched the
results at a value other than 2 GeV. Still, it is clear from Figures~\ref{fig:pert comp} and~\ref{fig:pert comp 0.0124} that a large range of momentum values exist where perturbation theory is expected to be reliable, and the precise matching
point will have little effect on the ultimate value of $a_\mu$. The matching point of 2 GeV is a natural choice, given the lattice scale in our study.
This comparison is a non-trivial check of the
lattice calculation of the vacuum polarization; in particular it indicates the correct number of physical quark degrees of freedom are accurately accounted for when using staggered fermions.

\begin{figure}[ht]
\begin{center}
\includegraphics[width=6in]{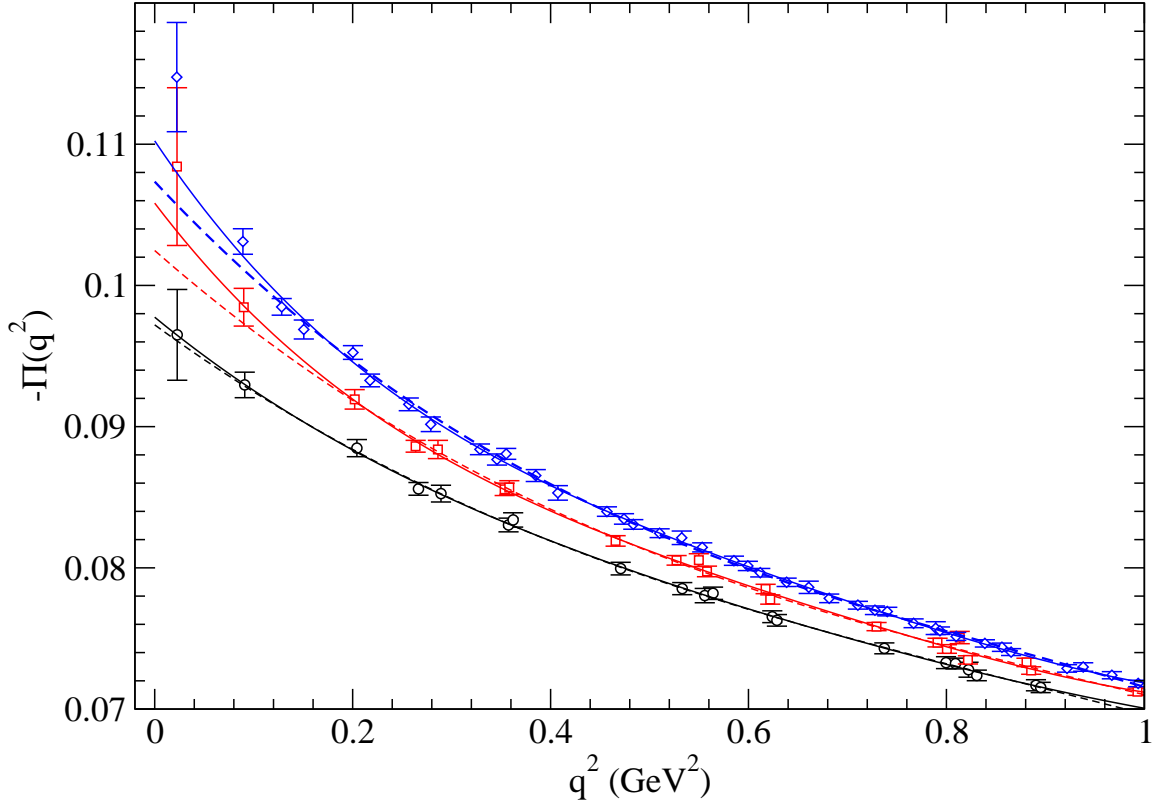}
\caption{Cubic (dashed) and quartic (solid) fits to $-\Pi(\hat q^2)$ for 
$am_l=0.0031$ (diamonds), $0.0062$ (squares), and 0.0124 (circles). The strange quark mass is fixed to 0.031 in each case.}
\label{fig:polyfits}
\end{center}
\end{figure}

For the low energy regime, we have tried several different fits.
In every case the fits are uncorrelated but performed under the single elimination jackknife procedure described above.  
Following previous calculations of the muon \gtwo\ \cite{g-2_1,g-2_2,g-2_3}, we try various polynomial fits, up to quartic order in $\hat q^2$. The general form for these fits is
\begin{equation}
	f(x) = \sum_{m=0}^n b_m x^m \, ,
\end{equation}
with $x = \hat q^2$, and $n$ ranging from 1 to 4. We will refer to these fits individually as ``Fit $n$.'' For larger $n$
these fits have adequate values of $\chi^2$, but tend to undershoot the lowest momentum points except for $n=4$. For $n<3$, the fits have very poor $\chi^2$ and do not fit the low-$\hat q^2$ reqion well. In Figure~\ref{fig:polyfits}, we display the cubic and quartic fits for each light quark mass. In Table~\ref{tab:polyfits}, we show the results for the fitting parameters for cubic and quartic fits at each of the three different quark masses. Additionally, we use these results to determine the $\hat q^2<1$GeV$^2$ contribution to the leading hadronic contribution to the muon \gtwo, and give these results in Table~\ref{tab:g-2results}. We will leave the discusion of these results to the next section.

More interesting are the fits using the \schpt\ vector resonance expressions for the vacuum polarization. We have performed three different fits:
\begin{description}
\item[Fit A:] Only the tree-level vector contribution given in \eq{rho_tree_level}.
\item[Fit B:] \eqs{Pi_pion_1taste}{rho_tree_level} which include the tree-level $\rho$ and the one-loop pion and kaon terms, but \textit{not} the one-loop terms in \eq{one_loop_rho}.
\item[Fit C:] The full one-loop expression, but absorbing
the one-loop $\rho$ analytic term into $f_V$.
\end{description}
Note that the last two fits give identical results for $a_\mu$ since
the pion and kaon logarithms and the NLO analytic term just serve to re-scale the (tree level) value of $f_V$ as the momentum dependence is identical to the tree-level expression.\footnote{Note this would not be possible if the vector masses were not identical at this order, since then there would be a nontrivial $q^2$ dependence.} In other words, the overall factor multiplying the $(p^2+m_V^2)^{-1}$ term stays the same, only the value of $f_V$ changes. Were we to be more precise, we would calculate the complete one-loop corrections to $f_V$ up to and including $
{\cal O}(q^4)$ terms and include these in the fitting functions. As our goal is to extract a value for the muon \gtwo, such a calculation, which for consistency also entails two loop diagrams in the pion sector, is beyond the scope of this paper.

\begin{figure}[ht]
\begin{center}
\includegraphics[width=6in]{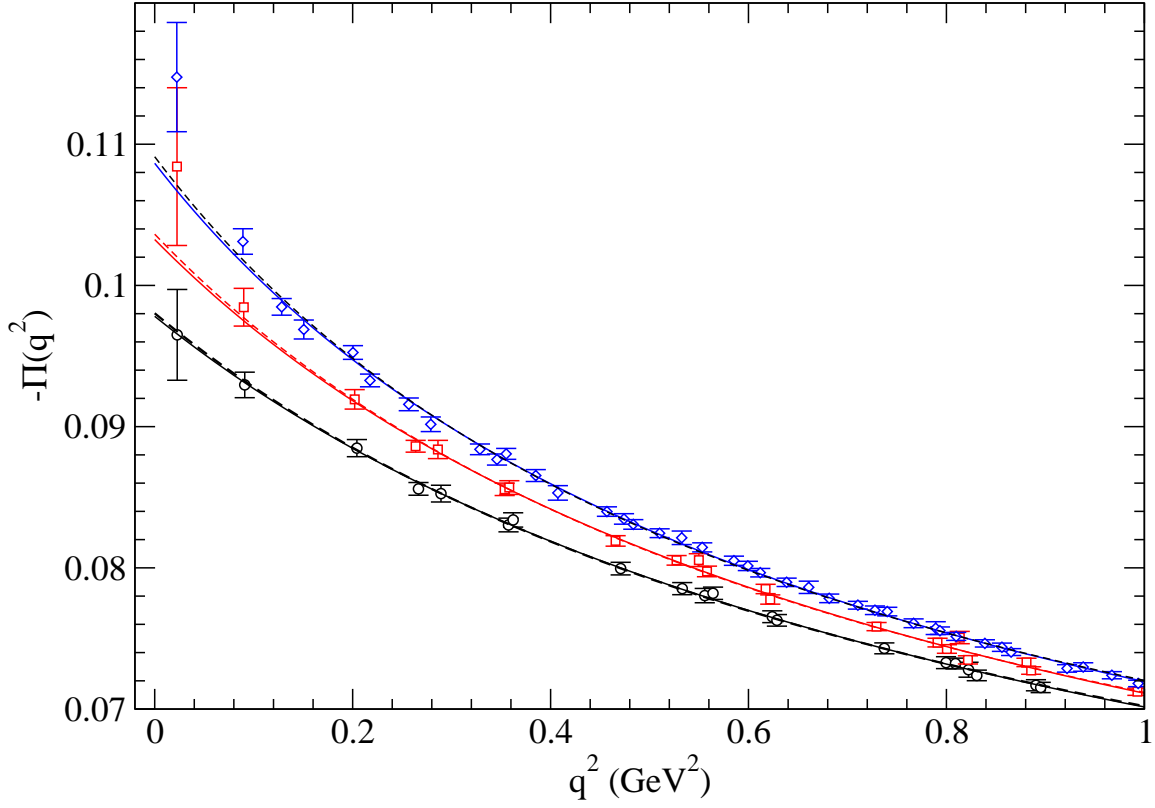}
\caption{\schpt\ fits to $\Pi(\hat q^2)$ for the three light 
masses, $am_l=0.0031$ (diamonds), $0.0062$ (squares), and 0.0124 (circles). The strange quark mass is fixed to 0.031 in each case. The solid lines correspond to Fit B, the dashed to Fit A, as described in the text.}
\label{fig:schptfits}
\end{center}
\end{figure}

The results from these fits with the maximum value of $q^2 =1$ GeV$^2$ are shown in Table~\ref{tab:schptfits}. The values for the decay constant change slightly from Fit
A to Fit B when adding in the pions, and there is some tendency for the decay constant to decrease with quark mass. In other words, the one-loop corrections from the pion sector have only a small influence. 
Fit C shows the tree-level value of the decay constant changes dramatically due to the one-loop corrections in \eq{one_loop_rho}.  
The fits are shown along with the data in Figure~\ref{fig:schptfits}. We take for the chiral scale $\Lambda =1$ GeV in Fits B and C.

The effect of decreasing the maximum value of $\hat q^2$ in the fit to 0.5 GeV$^2$
is small. The central value of $f_V$ changes by less than one percent, 
about two percent, and less than one percent for $m_l=0.0031$, 0.0062, and 0.0124, respectively. Raising the maximum to 1.5 GeV$^2$ induces larger changes, but the $\chi^2$ values of the fits are significantly worse. In all cases the vector masses are fixed to their 
nominal values from \cite{MILC_SPECTRUM}. The polynomial fits are less stable, with very large errors for the smaller range, and significant undershooting of the data for the largest range.
For these reasons we quote values of \amulo\ using the ``best" fit range,
$0\le q^2 \le 1.0$ GeV$^2$. 

\begin{figure}[ht]
\begin{center}
\includegraphics[width=5in,angle=270]{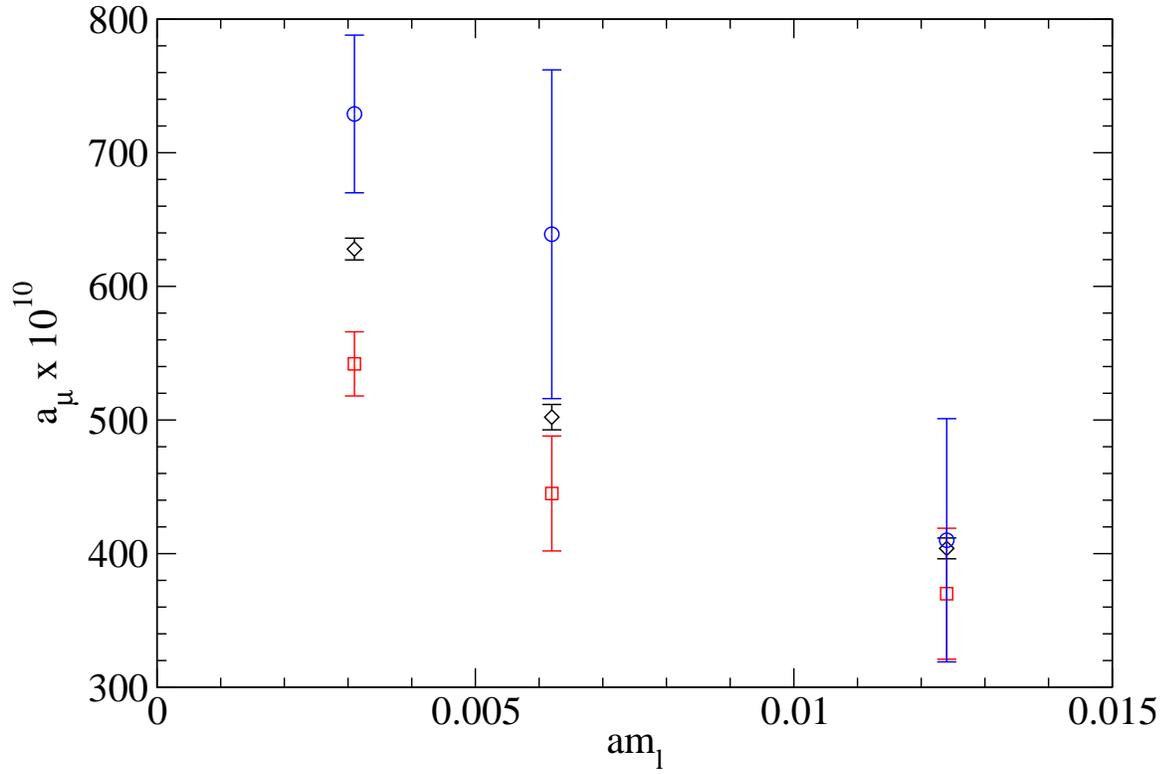}
\caption{The hadronic contribution to the muon anomalous magnetic moment from polynomial and \schpt\ fits to the vacuum polarization.
The statistical uncertainties for the cubic (squares) and quartic (circle)
fits are much larger than the \schpt\, one.}
\label{fig:g-2 all}
\end{center}
\end{figure}

The values for \amulo\ resulting from the above fits are listed in Table~\ref{tab:g-2results} and displayed in Figure~\ref{fig:g-2 all}, with only statistical errors shown. First, for the polynomial fits, we see a dramatic rise in \amulo\ as we decrease the quark mass, and also as we increase the order of the polynomial;
they are not stable in this sense. 
This is indicative of the calculation in general: the value of $\amulo$ is quite sensitive to the low momentum region, and hence the fit in this
region, due to the nature of \eq{kernel} and the smallness of the muon mass.
The low $\hat q^2$ region is fit better as the order of the polynomial increases which increases the value of \amulo\, but the errors on the fitted parameters increase such that the values for \amulo\ also have large errors. Thus, it is preferable to use one of the physically motivated fitting functions. For Fits A and B, we note there is little difference in the final result for \amulo, as expected from the fit results themselves. 
As mentioned in the last section, there is no difference in \amulo\ for Fits B and C since the one-loop corrections in \eq{one_loop_rho} only rescale the tree-level value of $f_V$. The statistical errors on \amulo\ for the \schpt\ fits are much smaller than the polynomial ones, so we use the former fits from now on to quote our best values.

From Figure~\ref{fig:schptfits} we see that the fits tend to undershoot the lattice calculation of $-\Pi(\hat q^2)$ for the lowest momenta for the smallest two quark masses, though within roughly a standard deviation. As for the polynomial fits, even small changes in the fits in this region lead to
large changes in \amulo . This undershooting behavior could represent real physics, or simply statistical and systematic errors. Certainly, the values of  $-\Pi(\hat q^2)$ at the smallest values of $\hat q^2$ are the most difficult to calculate, so the latter is more likely to be the case. The good fits obtained using chiral perturbation theory and the precisely measured meson masses from \cite{MILC_SPECTRUM},
over a wide range of momentum, also back up this explanation. Still, the possibility that we have not accounted for an effect due to small quark mass and momentum remains, and must be further investigated in future calculations. In such calculations it is important to reduce the statistical error on these points as much as possible. One way to do
this is to use a momentum source for each of the lowest momenta. This should have smaller errors than the point (-spilt) source used here, but is more complicated to implement and requires a separate propagator calculation for each momentum. Once the statistical errors
on the very low $\hat q^2$ region are reduced, one can begin to investigate systematics to tell whether the excess is an actual physical effect (which could increase the value of \amulo\ significantly). 

In a similar vein, we should also check the numerics of our calculation.
At the heart of the calculation is the quark propagator computation which is performed using the conjugate gradient algorithm to invert the lattice Dirac operator (matrix). The accuracy of the inversion is controlled by stopping the iterative solution found by the conjugate gradient algorithm when the norm of the so-called residual vector reaches some small size. The stopping criteria used here are given in Table~\ref{tab:latt_params} and are comparable in size to those used in other current simulations. They are also much smaller than the criteria used to generate the ensembles \cite{MILC_SPECTRUM}. One test was run on a single $40^3$, $m_l=0.0031$, lattice to check whether the stopping criteria was adequate. We calculated the vacuum polarization using three stopping criteria, $10^{-5}$, $10^{-6}$, and $10^{-7}$, and compared the results. The difference in last two calculations was always less than 0.1 \% except for two cases for the lowest momenta
where the difference was still less than 1\%, 
while the difference between the first two was always less than 1\%, except for a handful of cases including one for the lowest momenta where the difference was neary 10\%. Thus we chose to use a stopping criteria of $10^{-6}$ in this case (for all others we used $10^{-8}$), and we expect in all cases that the error in our results for the vacuum polarization due to a non-converged propagator solution is much smaller than the quoted statistical error.

The other possible problem with the numerics is the numerical precision of the calculation. In all cases the global sums in the quark propagator calculations were done in double precision and the rest of the calculation done in single (32-bit) precision, including the fourier transform of the vacuum polarization. This was also the procedure used in \cite{MILC_SPECTRUM}, although in that calculation they did a comprehensive check for systematic errors arising from the use of single precision arithmetic (see Table VIII of \cite{MILC_SPECTRUM}). Any differences due to precision were much smaller than quoted statistical errors, or even statistical errors that might be achieved by
a ``reasonable lengthening" of their run. 
 
 Thus for now we attribute the excess of the vacuum polarization over the fit results at the lowest one or two momenta for the smallest two quark masses to statistics and possible systematics (but not numerics), and rely on the fits to chiral perturbation theory to provide an accurate determination of the anomalous magnetic moment. We also note that the chiral perturbation theory fits for all quark masses are very stable if we leave out the lowest two momenta, as would be expected from the relative size of the statistical errors on these points. 
 
\begin{figure}[ht]
\begin{center}
\includegraphics[width=6in]{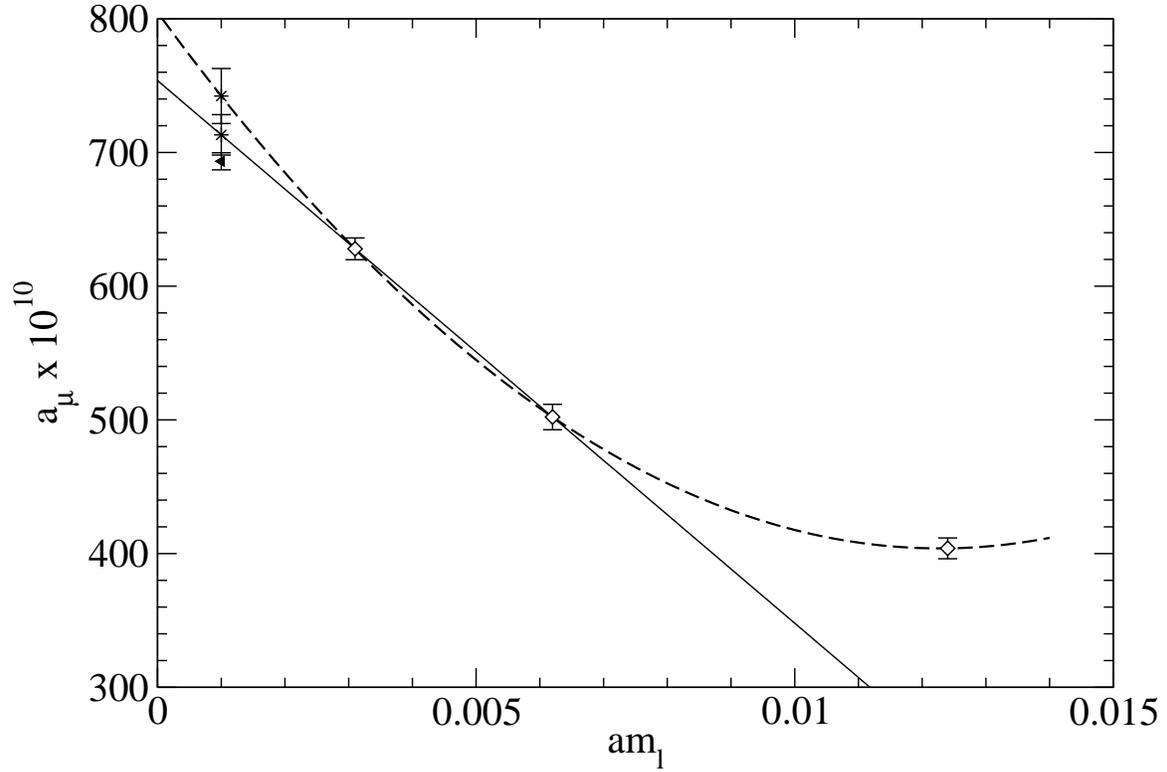}
\caption{Linear and quadratic extrapolations (bursts) 
of the hadronic contribution to the muon anomalous magnetic moment
to the physical point ($am_l\approx 0.001$). The left triangle is the currently accepted value, computed from the experimental cross-section for  $e^+e^-\to$ hadrons \cite{g2REVIEW,Davier:2004gb}.}
\label{fig:g-2extrap}
\end{center}
\end{figure}

We need to extrapolate the values of $\amulo$ at fixed quark mass to the physical point (which, in lattice units, is a light mass of about $am_l\approx0.001$ \cite{MILC_SPECTRUM,MILC_CHIRAL_FITS}). There are several ways to do this. The simplest way is to extrapolate \amulo\ in the light quark mass. This is shown in Figure~\ref{fig:g-2extrap}, with two different 
``phenomenological" fits. One is a linear fit in the lightest two masses, and the other is a quadratic fit to all three. The value from the
linear fit is 
\begin{equation}
a_\mu^{{\rm HLO},q^2\le1} = (713 \pm 15 )\times 10^{-10}
\end{equation}
while the quadratic gives 
\begin{equation}
a_\mu^{{\rm HLO},q^2\le1} = (742 \pm 21)\times 10^{-10},
\end{equation}
where the errors are statistical. Of course, there is no guarantee that (other) non-linear behavior does not set in as the two pion threshold is crossed, which we discuss below.

\begin{figure}[ht]
\begin{center}
\includegraphics[width=6in]{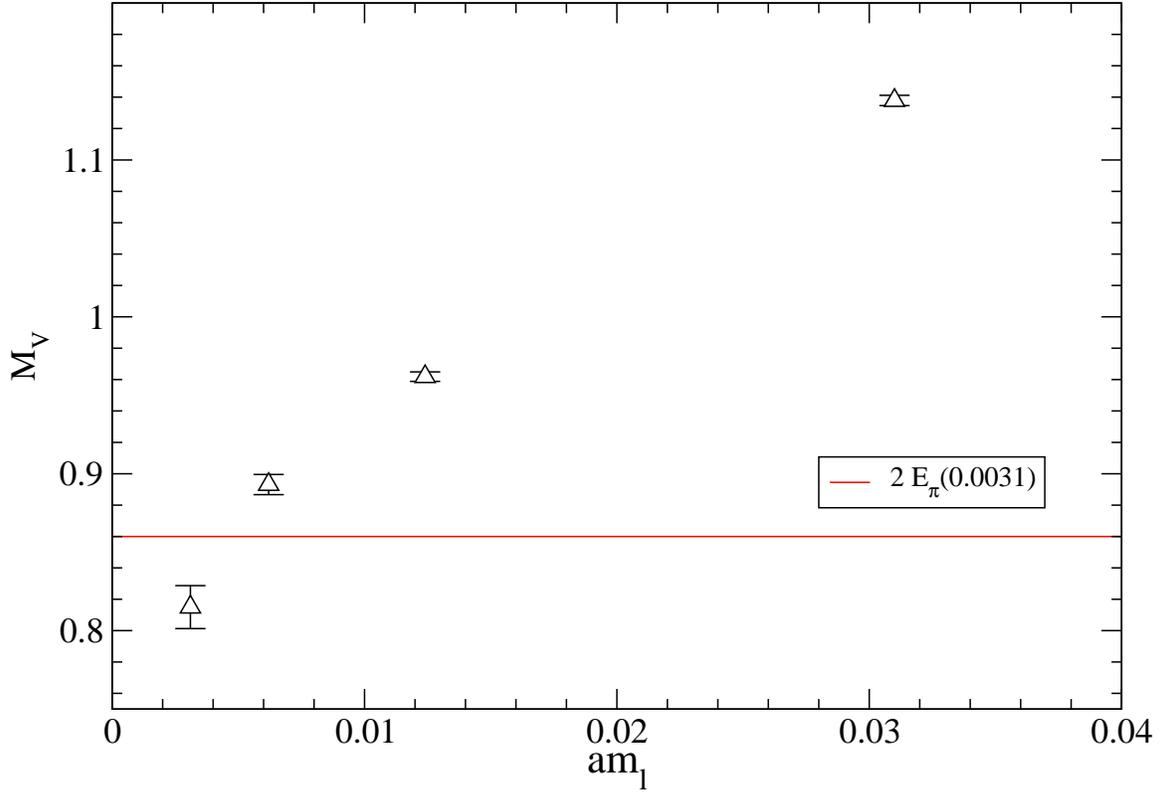}
\caption{The vector meson mass (GeV). Non-linearity is visible for the 
lightest quark mass studied here, and perhaps even the second
lightest mass. The horizontal line depicts the two pion energy for the
largest lattice and smallest quark mass in our study.
The mass values are from the MILC collaboration, \cite{MILC_SPECTRUM} and \cite{DOUG} ($am_l=0.0031$).}
\label{fig:vector mass}
\end{center}
\end{figure}

The more correct method to reach the physical value is to fit
all the lattice data simultaneously, and extrapolate the vector decay constant and  vector and pion masses to the physical light quark mass, with a single mass- and momentum- independent constant. 
The difficulty lies in the fact that such a fit requires computing the effects of the two pion threshold on the vector masses and decay constants, a clearly formidable task as the $\rho$ meson is unstable
for physical quark masses. In our study,
on the $40^3$ lattice with $am_l=0.0031$, the vector meson is just below the two pion threshold; however it shows clear non-linearity in the quark mass (see Figure~\ref{fig:vector mass}). Lacking knowledge of two pion effects in chiral perturbation theory to the specified order at which we are working,  we could instead simply develop smooth parameterizations of the masses (for example, see ~\cite{Armour:2005mk}) and decay constants and use them in our chiral perturbation theory formulae. But such a complicated procedure is not likely to yield more accurate results from this study than the simple extrapolation we have used.

Using the continuum three-loop perturbation theory result matched to the $am_l=0.0031$ lattice calculation at 2 GeV as described above, 
we find
\begin{eqnarray}
a_\mu^{{\rm HLO},q^2>1} &=& 6 \times 10^{-10}
\end{eqnarray}
for the high energy region, $q^2>1$ GeV$^2$. There is some small error in this result made by running down to 1 GeV$^2$ instead of some higher value more compatible with the lattice calculation as seen in Figure~\ref{fig:pert comp}. However, since the contribution to \amulo\ is already so small starting at 1 GeV$^2$ and the difference between the perturbative curve and the lattice data is small even at 1 GeV$^2$, we did not bother to fit the lattice data with a stepwise procedure up to such a high momentum. Note also that the perturbative curve is insensitive to the value of the light quark mass, so we did not attempt to extrapolate this contribution to the physical light quark mass.

Adding the contributions from high and low momentum regions, the total value of the lowest order hadronic contribution to \gtwo\ is
\begin{equation}
\amulo = (721 \pm 15)\times 10^{-10} 
\end{equation}
for the linear fit, and
\begin{equation}
\amulo = (748 \pm 21)\times 10^{-10},
\end{equation}
for the quadratic.
Our result is larger than the currently accepted theoretical numbers
\cite{g2REVIEW,Davier:2004gb}, 692.4 (5.9) (2.4) $\times 10^{-10}$ and 711.0 (5.0)(0.8)(2.8) $\times 10^{-10}$ , which use only $e^+e^-$ data and include $\tau$ decay data, respectively. Considering our statistical errors and the systematic uncertainties described below, the calculations should be viewed as compatible for now.

\section{Discussion}\label{sec:disc}

This calculation of \amulo\ has focused on a detailed study of precisely fitting the lattice computed vacuum polarization to obtain the value of \gtwo\ for several fixed quark masses. This has up to now only been done at a single lattice spacing, and as such, there are still discretization effects that could arise that we have not taken into account. Simulations using the coarser MILC lattices are underway \cite{AubinBlum}, and new lattices with a smaller lattice spacing than the 0.086 fm used here are being generated by the MILC collaboration, so the continuum limit of the vacuum polarization can be taken. One way to gauge the size of the non-zero lattice spacing errors in our
calculation, without actually repeating the calculation at another lattice spacing, is to use the naive continuum momenta $2\pi n /L$ everywhere in the analysis instead of the sin function ($2\sin{\pi n /L}$ appears in Ward-Takahashi identity so is the correct form to use). The resulting difference in
\amulo\ is within the statistical variation of the previous result.
Other evidence that non-zero lattice spacing errors are under control is the agreement between the lattice calculation of $\Pi(\hat q^2)$ and continuum perturbation theory, as already discussed in the previous section.

We also have not explicitly taken into account possible finite volume effects. The quark masses in these configurations are small enough to have small but not negligible finite volume corrections \cite{MILC_SPECTRUM,MILC_CHIRAL_FITS}. Most likely, however, these corrections are small compared with the statistical errors that exist in the data. In terms of our model used to fit the data, the dominant contribution is from a tree-level photon-$\rho$-photon diagram, where there will be no corrections coming from finite volume (other than the finite volume correction to the vector mass itself which is expected to be quite small). Finite volume corrections first appear in the one-loop diagrams and thus will be suppressed relative to the dominant contribution. 

Another source of systematic error lies in the mismatch of the fixed value of the strange quark mass on these lattices and the physical strange quark mass which is actually about 10 percent smaller. Although we have assumed that the effect of this mismatch is negligible, this may not be the case. We can see from the lattice data that there is significant quark mass dependence for the vacuum polarization for low values of momenta. Note that correcting for this
effect will tend to increase the value of $\amulo$.

Finally, as discussed already, the extrapolation to the physical light quark mass is not fully under control. Even though the extrapolations shown in Figure~\ref{fig:g-2extrap} appear sensible and are not large in the quark mass, the physical vector particles are unstable with non-zero widths, so  these effects should be properly included in our analysis of resonance chiral perturbation theory. Such a calculation was beyond the scope of the present study as a consistent treatment of the pion and vector sectors entails higher order calculations. 

\begin{figure}[ht]
\begin{center}
\includegraphics[width=6in]{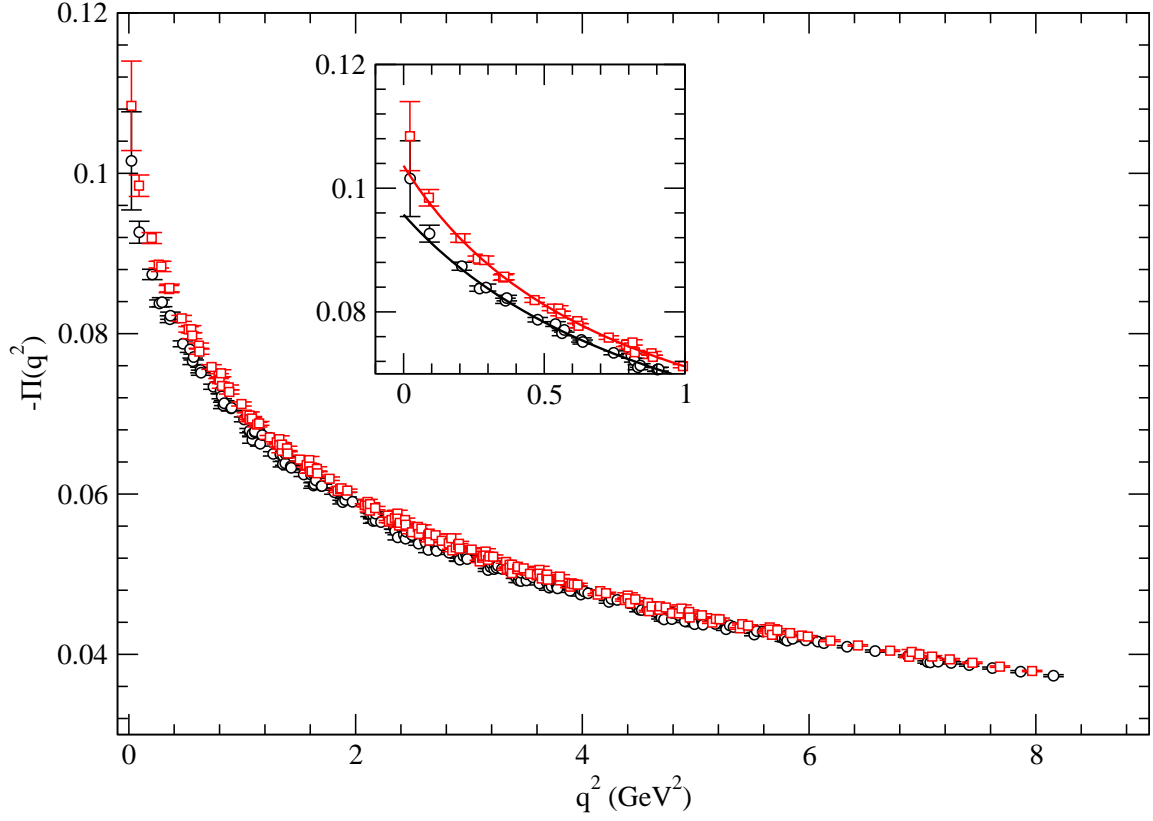}
\caption{The vacuum polarization computed on the quenched ensemble (circles) compared to the
2+1 flavor result. The light quark mass in each case is 0.0062, and
the quenched gauge coupling was tuned to match lattice spacings.
The solid lines are fits to \eq{rho_tree_level}.}
\label{fig:quenched comparison}
\end{center}
\end{figure}

In this context it interesting to discuss the effects due to quenching since it is in this approximation that the vector particles are stable. In Figure~\ref{fig:quenched comparison} we show the quenched value of  $-\Pi(q^2)$ computed on an ensemble of lattices \cite{MILC_SPECTRUM} with the same lattice spacing and volume as the $am_l=0.0062$, 2+1 flavor ensemble. The
quenched and 2+1 flavor results agree at large $\hat q^2$, but a noticeable deficit occurs for the quenched case as $\hat q^2\to 0$. The resulting value of \amulo\ is significantly lower (see Table~\ref{tab:g-2results}), consistent with previous quenched calculations\cite{g-2_1,QCDSF}.
Presumably this effect is caused by the sea quarks. 

Even though the $\rho$
mass is always below the two pion threshold in our calculation\footnote{The pions are required to have relative momentum, so the lowest pion energy is roughly $\sqrt{(2\pi n/L)^2+m_\pi^2} $}, these two pion states contribute to the vacuum polarization, albeit with larger invariant mass, and on the lattice, a discrete spectrum. In our \schpt\ analysis two pion effects come only from the direct coupling to the photon; the two pion coupling to vector mesons is ${\cal O}(p^4)$ and was not included. Because this coupling is higher order, we also expect that the effect of the non-zero width of the vector mesons on \amulo\ is suppressed.
The difference in the two cases is reflected, in the fits, through the vector mass itself which is significantly larger in the quenched case \cite{MILC_SPECTRUM}\footnote{In \cite{MILC_SPECTRUM} only light quark masses 0.015 and 0.03 were computed in the quenched case, so we simply extrapolated linearly in $am_l$ to obtain the value of the vector mass at 0.0062.} and the decay constant which is also larger
(see Table~\ref{tab:schptfits}). 
A more complete understanding of these effects is wanting, and must await both simulations at smaller quark masses
where the vectors are unstable and higher order calculations in chiral perturbation theory, some of which have been done already in continuum chiral perturbation theory \cite{Rosell:2004mn}. 
We conclude this dicussion by emphasizing that there is a significant effect, clearly visible in the lattice calculations (see Figure~\ref{fig:quenched comparison}), from unquenching which tends to increase the value of \amulo.

\section{CONCLUSION}
\label{sec:conc}

In this study we presented a value of the lowest order hadronic contribution to the muon anomalous magnetic moment computed in 2+1 flavor QCD. The statistical errors on our results at fixed light quark mass are small,
possibly underestimated, and comparable with the errors on the accepted value of these contributions computed using the experimentally measured total cross sections for $e^+e^-$ annihilation
to hadrons and tau lepton decay to hadrons\cite{g2REVIEW}. The central value found by extrapolation to the physical light quark mass 
is slightly larger than this value, with statistical errors about 2-3 times larger. Within the systematic uncertainties in our calculation which were discussed in detail in the previous sections, the values should be seen as compatible.
 
First, there is some uncertainty in our fitting procedure. Polynomial fits require a high degree to fit the low momentum region of the vacuum polarization so that the statistical errors on the parameters become large. The results are also somewhat unstable to changes in the fit (momentum) range. 

A more precise and stable fitting procedure was found by appealing to chiral perturbation theory, including lattice spacing effects. 
To our knowledge, this is the first time the hadronic vacuum polarization computed in chiral perturbation theory has been given in the literature. While the (staggered) chiral perturbation theory formula depends only on one (unknown) low energy constant and thus offers a useful fit function, we found that it does not represent the lattice calculation; to fit the data well requires the inclusion of the vector particles as well, through resonance chiral perturbation theory. This was already indicated by quenched calculations. The additional free parameter is the vector decay constant which shows modest quark mass dependence at tree level and large effects due to one-loop pion and kaon tadpole graphs.
The coupling of the vector meson to two pions which contributes to both mass renormalization and the non-zero width of the $\rho$ meson gives rise to a one-loop graph which is however ${\cal O}(p^4)$, and hence was not included in this work since a consistent treatment in the pion sector would require a two-loop computation.
We also emphasize that a direct comparison to a ``matched" quenched calculation showed a significant increase in the 2+1 flavor vacuum polarization at low momenta, presumably due to sea quark effects.
 
The vacuum polarization computed on the lattice was seen to be compatible with continuum weak-coupling perturbation theory over a wide range of momenta, between roughly 3 and 8 GeV$^2$. This indicates that lattice spacing errors may be under control in this study.
 
We have used the fits of the vacuum polarization to chiral perturbation theory to quote our final results. Remaining fitting uncertainties can be reduced in future calculations by reducing the statistical errors on the
smallest momentum values of $\Pi(q^2)$ and by including the ${\cal O}(p^4)$ terms in the chiral perturbation theory analysis. The size of the statistical error can be further reduced by increasing the ensemble size, computing with momentum sources, and computing at smaller valence quark mass. These require modest increases of computational resources compared to those used here. Reducing the sea quark mass is obviously very costly, but not
unforeseeable.

There are additional, as yet unquantified, systematic errors in our calculation, notably effects due to omission of disconnected quark diagrams giving rise to flavor singlet contributions, non-zero lattice spacing and unphysical quark masses. The latter were discussed at some length in the last section. The main point is that the two pion threshold may cause non-linearities in the light quark mass extrapolation which can be addressed by simulating below the threshold or, at least, by including higher order terms in the chiral expansion. It is also crucial to take the continuum limit and compute the vacuum polarization with different lattice fermions in light of the current understanding of the non-local nature of staggered fermions at non-zero lattice spacing.
While these effects will be addressed in future calculations, the current study of the muon \gtwo\ on the lattice in 2+1 flavor QCD appears quite encouraging.

\section*{Acknowledgements}

TB thanks Bill Marciano, Michael Ramsey-Musolf, and Doug Toussaint for interesting and helpful conversations. CA thanks Claude Bernard and Norman Christ for useful discussions. We especially acknowledge the MILC collaboration for providing the lattice ensembles necessary to carry out this work. The computer code used in our study is based on the freely available MILC QCD code, versions 6 and 7.
We thank the US Department of Energy (DOE) for support under an  Outstanding Junior Investigator award, No. DE-FG02-92ER40716 (TB) and grant No. DE-FG02-92ER40699 (CA).
This research was supported by the RIKEN BNL Research
Center (RBRC) (TB). We thank the DOE's NERSC Center, the RBRC, Brookhaven National Lab,
and Columbia University for providing generous supercomputer resources and support on the IBM SP2 and QCDOC machines, respectively.

\bibliography{references}

\vfill\eject

\begin{table}[ht]
\caption{Parameters for the MILC lattices and propagator calculations  in this study. ``c.g. stop res." is the stopping criterion, used for the residual vector in the conjugate gradient subroutine of the MILC code, for the valence quark propagator. The stopping criteria used to generate the lattices are given in Table I of \cite{MILC_SPECTRUM}.
In the ``confs." column total number of configurations are given for both propagator source time slices, $t=0$ and 48. $\beta=10/g^2$ is the bare gauge coupling for the tadpole-improved Symanzik gauge action used to generate the lattice ensembles \cite{MILC_SPECTRUM}. 
The $\pi$, $K$, and vector meson masses used in this study were taken from \cite{MILC_SPECTRUM} and \cite{DOUG} ($am_l=0.0031$).}
\begin{center}
\begin{tabular}{ccccccccccc}\hline
$a$ (fm) & size & $\beta$  & $am_l$ & $am_s$ & $am_{\rm val}$ & c.g. stop res. & confs. &$ am_\pi$& $ am_K$ & $ am_V$ 
\\\hline
0.086(2) & $28^3\times 96$ & 8.40 & - & - & 0.031 & 10$^{-8}$  & 84/54 \\
0.086(2) & $28^3\times 96$ & 8.40 & - & - & 0.0062 & 10$^{-8}$  & 111/54\\
0.086(2) & $28^3\times 96$ & 7.11 & 0.0124 & 0.031 & 0.031 & 10$^{-8}$  & 129/129  \\
0.086(2) & $28^3\times 96$ & 7.11 & 0.0124 & 0.031 & 0.0124 & 10$^{-8}$ &  129/129 & 0.20638 (18) & 0.27209 (18) & 0.4173 (13)\\
0.086(2) & $28^3\times 96$ & 7.09 & 0.0062 & 0.031 & 0.031 & 10$^{-8}$  & 42 \\
0.086(2) & $28^3\times 96$ & 7.09 & 0.0062 & 0.031 & 0.0062& 10$^{-8}$ & 249 &  0.14794 (19) & 0.25319 (19) & 0.3895 (28)  \\
0.086(2) & $40^3\times 96$ & 7.08 & 0.0031 & 0.031 & 0.031& 10$^{-8}$ & 51/51 \\
0.086(2) & $40^3\times 96$ & 7.08 & 0.0031 & 0.031 & 0.0031& 10$^{-6}$ & 161/153&  0.10521 (8)  & 0.24175 (15) & 0.357 (5) \\\hline\hline
\end{tabular}
\end{center}
\label{tab:latt_params}
\end{table}

\begin{table}[h]
\caption{Fit parameters for the cubic and quartic polynomials for the 2+1 flavor value of $-\Pi(\hat q^2)$. The fit range was taken to be
$0 \le \hat q^2\le 1$ GeV$^2$ in each case.
The jackknife estimates of the errors are statistical only, and the $\chi^2$ values are from uncorrelated fits.}
\label{tab:polyfits}
\begin{center}
\begin{tabular}{cccccccc}
\hline\hline
order & $am_l$    &  $b_0$&  $b_1$&  $b_2$&  $b_3$&  $b_4$ & $\chi^2$/dof
\\\hline
3 & 0.0124  & 0.0972(14)  & -0.051(7) & 0.038(13) & -0.014(7)  & --- & 3.5/16\\
4 & 0.0124  & 0.0978(20)   & -0.058(14)  & 0.062(39) & -0.049(45) & 0.017(19) & 3.4/15 \\ 
3 & 0.0062  & 0.1025 (14)  & -0.0615 (63) & 0.045 (10) & -0.015(5)  & --- & 19.3/7 \\
4 & 0.0062 & 0.1058 (27) & -0.092 (19) & 0.137(48) & -0.124(51) & 0.045(19) & 16.3/16\\ 
3 & 0.0031  &  0.1075 (9)  & -0.0754 (35) & 0.0636 (54)  & -0.0241 (27)& --- & 25.6/35\\
4 & 0.0031  & 0.1102(13)   & -0.1044(89) & 0.157(23)  & -0.142(26) & 0.051(10) & 17.4/34 \\ \hline
\end{tabular}
\end{center}
\end{table}%

\hskip -.5in
\begin{table}[ht]
\caption{Fit parameters for \schpt\ formulae for the 2+1 flavor value of $-\Pi(\hat q^2)$. The first row is for the quenched case discussed in the text. The fit range was taken to be
$0 \le \hat q^2\le 1$ GeV$^2$.
The jackknife estimates of the errors are statistical only, and the value of $\chi^2/$dof is from an uncorrelated fit. The meson masses were fixed to the values given in Table~\ref{tab:latt_params}. The staggered meson mass splittings used in Fit B and Fit C (not shown) are found in \cite{MILC_SPECTRUM}. }
\begin{center}
\hskip -.0in
\begin{tabular}{cccccccccc}\hline
$am_l$ & $f_V^{\rm Fit A}$ (MeV) &  $A^{\rm Fit A}$ & $\chi^2$/dof &
$f_V^{\rm Fit B}$ (MeV) &$A^{\rm Fit B}$ & $\chi^2$/dof & $f_V^{\rm Fit C}$ (MeV) &$A^{\rm Fit C}$ & $\chi^2$/dof \\\hline
0.0062 & 209.9 (2.0) & 0.0410 (6) & 23/18\\
0.0124  &  192.8 (1.8) &0.0445 (6) & 4.4/18 &  188.6 (2.0) & 0.0421 (7) & 4.0/18 &  117.2 (1.2) & 0.0420 (6) & 4.0/18 \\
0.0062 & 186.8 (1.7) & 0.0453 (5) & 18/19 &  181.6 (1.8) & 0.0422 (5) & 18 /19 & 115.7 (1.0) & 0.0422 (5) &  18/19 \\
0.0031  & 175.4 (1.1) & 0.0474 (3) & 28/37 & 169.9 (1.1)  & 0.0436 (3) & 25/37 &
111.5 (7) & 0.0436 (3) & 25/37 \\\hline
\end{tabular}
\end{center}
\label{tab:schptfits}
\end{table}

\begin{table}[ht]
\caption{Results for $\amulo\times 10^{10}$ for the various fits described in the text. Errors are jackknife estimates and statistical only. The quenched results correspond to light valence quark mass 0.0062 and strange valence quark mass 0.031.}
\label{tab:g-2results}
\begin{center}
\begin{tabular}{ccccc}
\hline
Fit  & quenched & $am_l=0.0124$ &  $am_l=0.0062$ & $am_l=0.0031$  \\\hline
Poly 3  & 381 (63) & 370(49) & 445(43) & 542(24) \\
Poly 4  & 588 (142) & 410(91)&  639(123) & 729(59)\\
A  &  366.6 (7.0) & 412.3 (7.8) & 516.0 (9.5) &  646.9 (8.1)  \\
B  & & 403.9 (7.8)  & 502.1 (9.5) & 628.0 (8.1) \\
C  &  & 403.9 (7.8) & 502.1 (9.5)  &  628.0 (8.1) \\
\hline
\end{tabular}
\end{center}
\end{table}%

\clearpage

\clearpage

\end{document}